\pdfoutput=1

\documentclass[sigconf]{acmart}

\usepackage{amsmath}

\usepackage{amssymb}
\usepackage{amsfonts}
\usepackage{algorithm}
\usepackage[noend]{algpseudocode}
\usepackage{graphicx}
\usepackage{textcomp}
\usepackage{xcolor}
\usepackage{caption}
\usepackage{subfigure}
\usepackage{booktabs}
\usepackage{tabularx}
\usepackage{hyperref}
\usepackage{multirow} 
\usepackage{svg}
\usepackage{float}
\def\BibTeX{{\rm B\kern-.05em{\sc i\kern-.025em b}\kern-.08em
    T\kern-.1667em\lower.7ex\hbox{E}\kern-.125emX}}
\AtBeginDocument{%
  \providecommand\BibTeX{{%
    Bib\TeX}}}

\copyrightyear{2024}
\acmYear{2024}
\setcopyright{rightsretained}
\acmConference[ICPP '24]{The 53rd International Conference on Parallel Processing}{August 12--15, 2024}{Gotland, Sweden}
\acmBooktitle{The 53rd International Conference on Parallel Processing (ICPP '24), August 12--15, 2024, Gotland, Sweden}
\acmPrice{}
\acmDOI{10.1145/3673038.3673142}
\acmISBN{979-8-4007-1793-2/24/08}

\begin{document}

\title{Bandwidth-Aware and Overlap-Weighted Compression for Communication-Efficient Federated Learning}



\author{
    Zichen Tang\textsuperscript{$\dag$}, 
    Junlin Huang\textsuperscript{$\dag$}, 
    Rudan Yan\textsuperscript{$\dag$}, 
    Yuxin Wang\textsuperscript{$\ddag$}, 
    Zhenheng Tang\textsuperscript{$\ddag$}\textsuperscript{\textasteriskcentered}, 
    Shaohuai Shi\textsuperscript{$\S$} \\
    Amelie Chi Zhou\textsuperscript{$\ddag$}, 
    Xiaowen Chu\textsuperscript{$\dag$}
    }
\authornote{Correspondence to: Zhenheng Tang and Xiaowen Chu.}
\affiliation{\institution{\textsuperscript{$\dag$}The Hong Kong University of Science and Technology (Guangzhou) \country{China} \\
\textsuperscript{$\ddag$}Hong Kong Baptist University \country{Hong Kong, China}
\textsuperscript{$\S$}Harbin Institute of Technology, Shenzhen \country{China}}
}
\email{
{ztangap, jhuang688, ryan987}@connect.hkust-gz.edu.cn,
}
\email{
{yxwang, zhtang, amelieczhou}@comp.hkbu.edu.hk, 
shaohuais@hit.edu.cn, 
xwchu@ust.hk
}

\renewcommand{\shortauthors}{Zichen Tang, Junlin Huang, Rudan Yan, et al.}

\begin{abstract}
Current data compression methods, such as sparsification in Federated Averaging (FedAvg), effectively enhance the communication efficiency of Federated Learning (FL). However, these methods encounter challenges such as the straggler problem and diminished model performance due to heterogeneous bandwidth and non-IID (Independently and Identically Distributed) data. To address these issues, we introduce a bandwidth-aware compression framework for FL, aimed at improving communication efficiency while mitigating the problems associated with non-IID data.
First, our strategy dynamically adjusts compression ratios according to bandwidth, enabling clients to upload their models at a close pace, thus exploiting the otherwise wasted time to transmit more data.
Second, we identify the non-overlapped pattern of retained parameters after compression, which results in diminished client update signals due to uniformly averaged weights. Based on this finding, we propose a parameter mask to adjust the client-averaging coefficients at the parameter level, thereby more closely approximating the original updates, and improving the training convergence under heterogeneous environments.
Our evaluations reveal that our method significantly boosts model accuracy, with a maximum improvement of 13\% over the uncompressed FedAvg. Moreover, it achieves a $3.37\times$ speedup in reaching the target accuracy compared to FedAvg with a Top-K compressor, demonstrating its effectiveness in accelerating convergence with compression.
The integration of common compression techniques into our framework further establishes its potential as a versatile foundation for future cross-device, communication-efficient FL research, addressing critical challenges in FL and advancing the field of distributed machine learning.
\end{abstract}




\maketitle
\section{Introduction}
Federated Learning (FL) has emerged as a promising machine learning paradigm that enables multiple parties to jointly train a model while keeping the training data samples decentralized without sharing. In the scenario of FL, each party (client) performs training on the local dataset and communicates model updates with a central server. This approach protects data privacy \cite{mcmahan2017communication,li2022federated} and reduces the need for central data storage, as only model updates are transmitted and aggregated. 

The main challenge along with FL is its communication bottleneck \cite{shah2021model,han2020adaptive,chen2021communication,konevcny2016federated,nguyen2024towards}. Numerous clients attempt to communicate local updates with the central server, making huge communication overheads bottleneck the training performance. To tackle the communication bottleneck, Federated Averaging (FedAvg) \cite{mcmahan2017communication} has been proposed that involves sampling only a fraction of clients to participate in model training and communication, and enabling more iterations of local computation, thus reducing communication costs. In each communication round, clients download the global model from the global server and perform several iterations of SGD. Local model updates are sent back to the global server for averaging, generating a new global model for the next round (Section~\ref{Section:FL}).

However, FL also presents certain limitations that have not been addressed by FedAvg, including data heterogeneity, i.e., non-independently and identically distributed (non-IID) data~\cite{kairouz2021advances,VHL}, and system heterogeneity such as varying network bandwidth~\cite{GossipFL}. The heterogeneous bandwidth\footnote{Given that clients may be distributed over a wide geographic area, such a scenario inherently leads to heterogeneous bandwidth across the network, which intensifies the straggler problem. } leads to the straggler problem in FedAvg where the synchronization setting necessitates waiting for the slowest client to finish transmitting the model update before the next communication round. This is especially significant in cross-device FL, where clients are edge devices such as IoT devices and portable electronics with unstable and constrained network connections. The disparity in bandwidth prolongs the whole training duration, leading to delayed model convergence and reduced efficiency. Therefore, it becomes imperative to design an advanced FL approach that ensures efficient and timely convergence under the heterogeneous bandwidth setting (Section~\ref{Section:BCRS}). 

Several variants have effectively enhanced FedAvg with model compression techniques to address the critical issue of communication overhead. By incorporating methods such as quantization and sparsification \cite{reisizadeh2020fedpaq, chen2021communication,shah2021model,han2020adaptive,jiang2022model,shi2020quantitative,shi2021towards}, those methods significantly reduce the size of the model updates that need to be transmitted.
While these compression algorithms prioritize communication efficiency, they do not consider data heterogeneity, which is a practical problem in real-world FL. Moreover, compression algorithms employing uniform compression ratios fail to account for bandwidth heterogeneity, thereby continuing to be susceptible to the straggler problem. Data heterogeneity concerned works alleviate the impact of data heterogeneity but impose extra communication burdens~\cite{karimireddy2020scaffold,luo2021no,tang2024fedimpro} on resource-constrained clients. These limitations underscore a broader issue in FL: \textbf{the challenge of designing algorithms that are communication efficient while effectively addressing data heterogeneity, ensuring both the robustness and accuracy of the global model} (Section~\ref{Section:OPWA}).

In this paper, we introduce a novel compressed FL framework to enhance communication efficiency under heavily heterogeneous data distribution. We develop Bandwidth-aware Compression Ratio Scheduling (BCRS) that dynamically adjusts the compression ratios and client-averaging coefficients based on bandwidth, enabling high-bandwidth clients to contribute more non-zero parameters in the averaging process, thus accelerating the global model convergence. Our research also uncovers unique distribution patterns of retained parameters after compression, which causes the diminishing significance of parameter updates retained infrequently due to the uniform averaging strategy. Leveraging this insight, we employ a parameter mask in Overlap-aware Parameter Weighted Averaging (OPWA) to compensate for inadequacies of FedAvg's uniform averaging strategy in aggregating model updates to expedite convergence. We conduct extensive comparative experiments on different datasets to demonstrate the robustness and improved accuracy of our algorithms. In our evaluations, our method demonstrates significant improvements in model accuracy, achieving a maximum increase of 13\% compared to uncompressed FedAvg. Additionally, it achieves a $2.02-3.37\times$ speedup in reaching the target accuracy compared to FedAvg with a Top-K compressor.
We also incorporate several commonly used compression techniques into our compressed FL framework, facilitating the execution of cross-device communication-efficient FL experiments in future research.

The key contributions of this work are listed as follows:
\vspace{-0.1cm}
\begin{itemize}
    \item We have developed a Bandwidth-aware Compression Ratio Scheduling (BCRS) algorithm that models the uplink communication time and automatically adjusts the compression ratios and client-averaging coefficients according to bandwidth conditions.
    \item We have discovered the heterogeneous distribution pattern of the retained parameters after compression and defined a new metric to quantify this parameter distribution. 
    \item We have introduced an innovative Overlap-aware Parameter Weighted Average (OPWA) algorithm that uses a parameter mask to adjust the averaging weights of parameters after magnitude pruning based on their occurrence frequency across clients. This novel averaging strategy is independent of compression algorithms and can be combined to enhance the performance of model compression.
    \item We have conducted extensive experiments on several datasets to robustly demonstrate the superior performance and effectiveness of our BCRS and OPWA algorithms.
\end{itemize}

\section{Related Work}
\label{Sec:relatedwork}
Client heterogeneity in FL encompasses both data and system heterogeneity, each posing distinct challenges to scalability and practical implementation. Subsequent subsections explore existing literature on these critical aspects. 
\subsection{Data Heterogeneity}
Data heterogeneity refers to the scenario in FL that each party’s local dataset cannot represent the overall distribution, making the data non-IID distributed. Although the basic framework FedAvg has been shown to achieve good performance empirically to overcome the data heterogeneity, it still fails to generalize convergence guarantee in even convex optimization settings \cite{li2019convergence,li2020federated}. 

Several works have stepped forward to provide theoretical convergence analysis under the non-IID setting. Adaptive optimization methods are employed in \cite{tang2024fedimpro,wu2023faster,reddi2020adaptive,wang2022communication} in response to the disparity of data distribution. FedProx \cite{li2020federated} offers a distinct approach by adding a proximal term to the clients’ local objectives, thus mitigating the mismatch between local and global optima. The methodologies can be generally categorized into feature calibration~\cite{VHL,luo2021no}, model customization \cite{liang2020think,t2020personalized,li2021ditto,collins2021exploiting,Dong_2023_CVPR,Dong_2023_ICCV}, multi-task learning \cite{marfoq2021federated,chen2018federated} and meta-learning \cite{fallah2020personalized}. 

\vspace{-0.3cm}
\subsection{Communication compression in FL}
Due to limited bandwidth in internet connections, the transmission between servers and clients has become an inherent bottleneck~\cite{tang2023fusionai,tang2020survey}, adversely affecting FL performance. Consequently, there is an urgent need for practical FL deployment to reduce communication overhead, especially in large language model scenarios~\cite{tang2023fusionai,Wang2023ReliableAE,Wang2024BurstGPTAR}. Sparsification has emerged as an effective method to decrease the number of parameters transmitted. Gradient Sparsification (GS) involves pruning model updates using magnitude-based or importance-based pruning \cite{molchanov2019importance,kirkpatrick2017overcoming,das2023beyond,shi2020quantitative,shi2021towards}. Studies \cite{jiang2018linear,wangni2018gradient,shi2019distributed} propose a periodic averaging GS strategy that randomly prunes a subset of gradients, allowing iteration over the entire gradient set within a few communication rounds.


Another direction in sparsification involves training personalized sparse models. \cite{bibikar2022federated,jiang2022model,qiu2022zerofl,GossipFL,tang2020communication,dong2024pruner} introduce a high level of sparsity in the local model training stage, effectively reducing the number of transmitted parameters. Works in \cite{nguyen2024towards,vogels2023communication,konevcny2016federated} used a low-rank method to train personalized sparse models. In this paper, we mainly consider generic FL, where all the clients share the same model structure.

Orthogonal to Sparsification, quantization emerges as another pivotal strategy to alleviate the communication bottleneck. This approach represents model updates in lower bits compared to the previous 32 or 64 bits, reducing the numerical precision. FedPAQ \cite{reisizadeh2020fedpaq} adopts a periodic averaging of the low-bit representation of local model updates to reduce communication frequency and overhead per round. \cite{gupta2022quantization} has made further advancements by refining quantization techniques. This work introduces a variant of Quantization-Aware Training (QAT) that is robust to multiple bit-widths, eliminating the need for retraining in the FL setting.

To the best of our knowledge, only a few studies consider solving both data heterogeneity and communication bottlenecks. 
Some works tackle data heterogeneity but increase communication burdens or degrade performance when paired with communication-efficient methods. Others prioritize communication efficiency but overlook the impact of data heterogeneity. In our approach, we do not introduce a new compression algorithm; instead, we propose a novel averaging weight adjustment strategy from both client and parameter levels, which can be integrated with existing sparsification techniques. This innovation strikes an intriguing balance between communication cost and model accuracy under the data heterogeneity setting.

\section{Preliminary}
\label{Sec:RELIMINARY}
\subsection{Definitions and Notations}
\label{Sec:notation}
For clarity and ease of understanding, the commonly used notations are summarized in Table~\ref{tab:main_notation}.

\begin{table}[ht]
\vspace{-0.3cm}
\centering
\caption{Main Notation.}
\label{tab:main_notation}
\begin{tabularx}{\columnwidth}{c|X} 
\hline 
Symbol & Description \\
\hline 
\hline
\( B_i \) & Bandwidth for the $i$-th client \\
\hline
\( L_i \) & Latency for the $i$-th client \\
\hline
\( N \) & Number of clients \\
\hline
\( C \) & The fraction of clients selected in each round \\
\hline
\( \mathcal{S}_t \) & The set of selected clients with size N*C in round $t$\\
\hline
\( E \) & Number of local epochs each client performs \\
\hline
\( M \) & parameter mask (same size as the model update) \\
\hline
\( \alpha \) & Hyperparameter: Server learning rate in averaging\\
\hline 
\( \gamma \) & Hyperparameter: Enlarge rate for specified parameter \\
\hline
\( \eta \) & Hyperparameter: Local learning rate \\
\hline
\( w_{it}^{\text{sparse}} \) & Sparsified model of the $i$-th client in round $t$ \\
\hline
\( T_{comm,i} \) & Communication time for the $i$-th client\\
\hline
\( T_{bench} \) & Compressed communication time of the slowest client\\
\hline
\( p_{i} \) & Averaging coefficient for the $i$-th client\\
\hline
\( CR \) & Compression ratio\\
\hline
\( f_i \) & Data frequency for the $i$-th client\\
\hline
\( V \) & Size of the transmitted model\\
\hline
\( \beta \) & Data heterogeneity level (Lower is more severe)\\
\hline
\hline
\end{tabularx}
\end{table}
\vspace{-0.4cm}
\subsection{Federated Learning}
\label{Section:FL}
\textbf{Federated Learning} is designed to cooperatively train a global model denoted by $w$ while circumventing the necessity to directly access the local data distributed among each client. Particularly, FL aims to minimize the objective of the global model $F(w)$:
\begin{equation} \label{fl_g}
    \min_{w} F(w) \triangleq \sum_{k=1}^{N} p_k F_{k}(w),
\end{equation}
where $N$ denotes the total number of clients, $p_k \geq 0$ is the averaging coefficient of the client $k$ such that $\sum_{k=1}^{N} p_k=1$, and $F_k(w)$ is the local objective measuring the local empirical risk defined as:
\begin{equation} \label{fl_l}
    F_{k}(w) \triangleq 
    \mathbb{E}_{(x,y) \sim \mathcal{P}_k(x,y)
    }
    \ell (f(x;w), y),
\end{equation}
with $\mathcal{P}_k$ representing the joint distribution of data in client $k$.
FedAvg stands out as a fundamental algorithm that efficiently aggregates model updates from multiple decentralized devices. In each communication round $t$, clients download the global model $w$ from the central server and perform $E$ epochs of stochastic gradient descent (SGD) on selected client set $\mathcal{S}_t$, where $E$ is a predefined constant and $\left| \mathcal{S}_t\right| = N \times C$ represents a small fraction $C$ of clients selected for round $t$. 
\begin{equation} \nonumber \label{l_ud}
    w_{k, j+1}^{t} \leftarrow w_{k, j}^{t} - \eta_{k, j} \nabla J_{k}(w_{k, j}^{t}), j = 0, 1, \cdots, E-1,
\end{equation}
where $w_{k, j}^{t}$ represents the $j$-th updates for the $k$-th client at round $t$, i.e., $w_{k, 0}^{t} = w^{t}$, and $\eta_{k, j}$ is the learning rate. 
At the end of round $t$, local model updates of selected clients are averaged by the central server, generating a new global model $w_{t+1}$ for the round $t+1$:
\begin{equation} \nonumber \label{avg}
    w^{t+1} \leftarrow \sum_{k \in \mathcal{S}^{t}} p_{k} w_{k, E-1}^{t},
    \ 
    p_{k} = \frac{n_{k}}{\sum_{i \in \mathcal{S}_t} n_{i}},
\end{equation}
with $n_k$ being the number of samples on the $k$-th client.

Despite its empirical success in non-IID settings, FedAvg still lacks a convergence guarantee for non-convex problems. Severe data heterogeneity can lead to the \textit{client shift} problem, where there is a mismatch between the global optima $w^{*}$ and local optima $w_i^{*}$, impacting the overall performance of the global model.

\begin{figure}[ht]
\vspace{-0.3cm}
\setlength{\abovedisplayskip}{-1pt}
    \subfigbottomskip=2pt
    \subfigcapskip=1pt
    \setlength{\abovecaptionskip}{0.cm}
  \centering
	\includegraphics[width=0.8\linewidth]{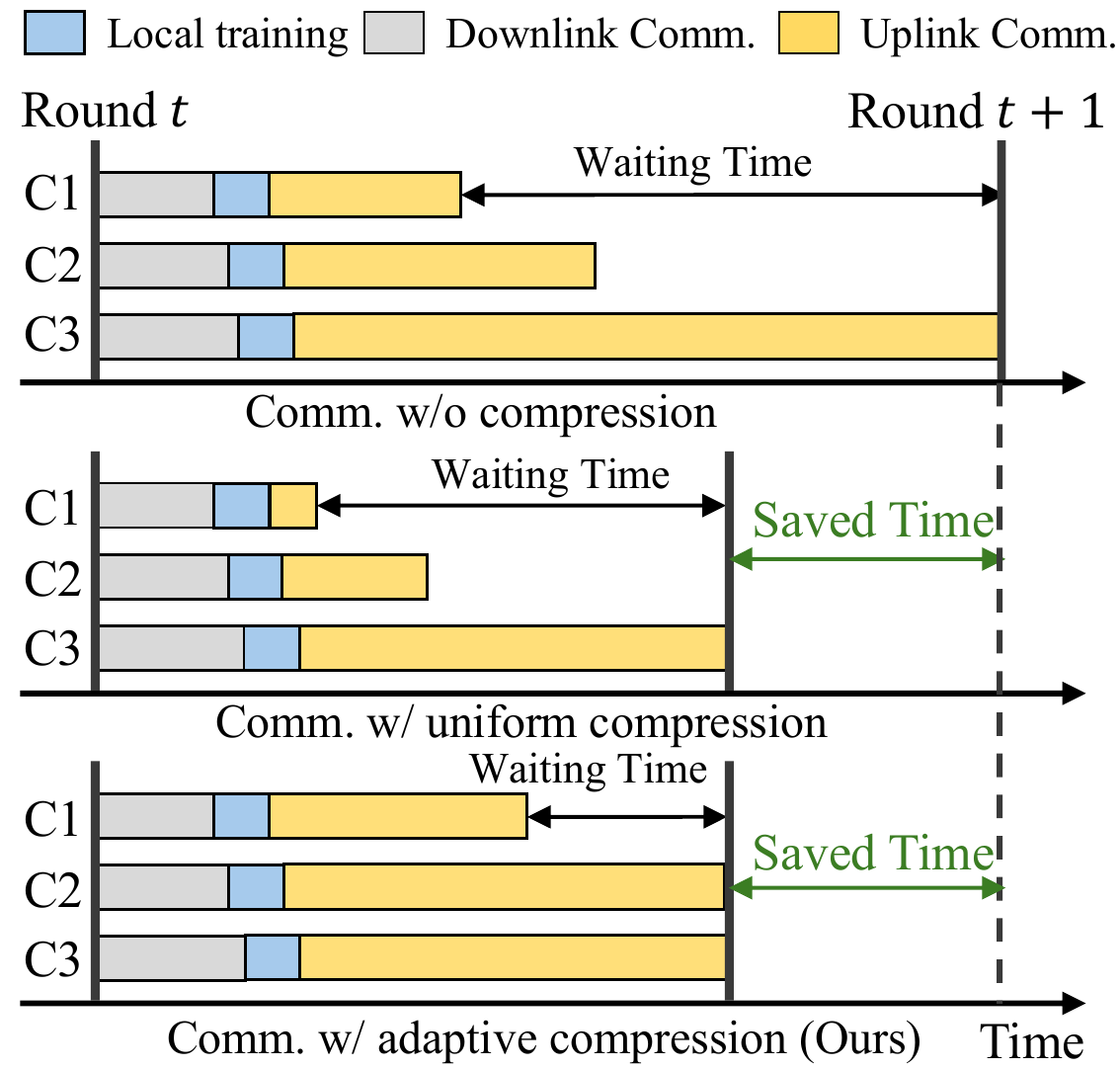}
	\caption{Timelines of different methods with FedAvg. Comm. represents communication, C1, C2, C3 represent three different clients. $B_1 > B_2 > B_3$ for these clients.}
	\label{fig:Straggler}
\vspace{-0.3cm}
\end{figure}

\begin{figure}[ht]
\vspace{-0.2cm}
\centering
\setlength{\abovedisplayskip}{-1pt}
    \setlength{\abovecaptionskip}{0.4cm}
	\includegraphics[width=0.7\linewidth]{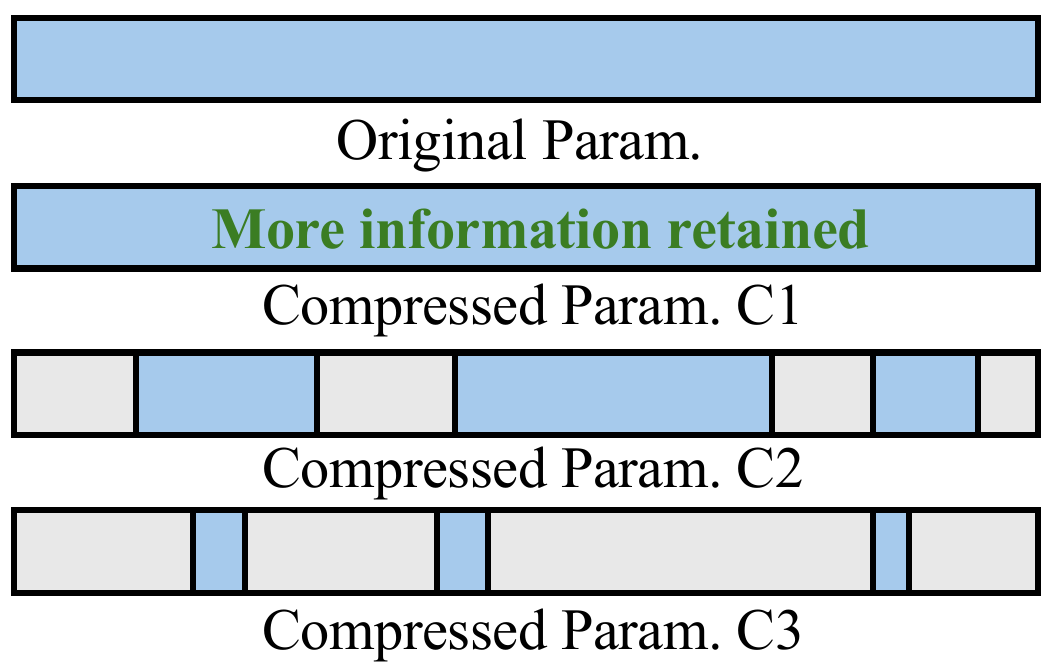}
	\caption{Adaptive communication ratios based on client bandwidth $B_1 > B_2 > B_3$. Such methods make clients 1 and 2 retain as much information as possible while guaranteeing the communication time will not be larger than the uniform compression.}
	\label{fig:adaptive-compression}
\vspace{-0.5cm}
\end{figure}

Although FedAvg employs synchronous SGD on a small fraction of selected clients to mimic scenarios where not all clients complete the designated computation, it does not fully address system heterogeneity. FedAvg still struggles with synchronization issues due to heterogeneous bandwidth. In the entire training process, clients with better network connections have to wait for the slower ones:
\begin{equation}
    T_{comm} = \max (T_{comm,i}) \quad \text{where} \quad i = 1, 2, \ldots, \left| S_t\right|
\end{equation}
This leads to the exacerbated straggler problem and inefficient utilization of connection resources, illustrated in the upper figure in Fig.~\ref{fig:Straggler}.
In our work, we model the communication time following the cost model in \cite{thakur2005optimization} to simulate real-world FL conditions. The definitions are given as follows: 
\begin{equation}
    T_{comm} = L + \frac{V}{B},
\end{equation}
where Latency $L$ is a time interval per message independent of message size. Bandwidth $B$ is the maximum rate of data transfer, typically measured in bits per second (bps). 
Details about the implementation are covered in Section~\ref{Section5B:Measurement_Bandwidth_latency}.
\subsection{FL with compressed communication}
\label{sec:flcompression}
Increasing the size of training datasets and expanding the parameter space in DNNs can effectively boost predictive performance across various applications. This escalation in model size directly contributes to heightened communication burdens, particularly in distributed environments like FL. Compressed FedAvg effectively mitigates the heavy communication overheads inherent in transmitting large and complex ML models, especially for edge devices whose bandwidths are significantly lower than intranet-based networks.
In FedAvg, the model updates are given by:
\[ \Delta w_t = \frac{1}{K} \sum_{k=1}^{K} \Delta w_{t,k} \]
where \( \Delta w_{t,k} \) represents the update from the \( k \)-th client at round \( t \), and \( K \) is the number of selected clients. Compression methods, such as sparsification, reduce the size of \( \Delta w_{t,k} \). For instance, sparsification can be represented as:
\vspace{-0.05cm}
\[ \Delta w_{t,k}^{sparse} = \text{Sparsify}(\Delta w_{t,k}) \]
\vspace{-0.05cm}
As shown in Fig.~\ref{fig:Straggler} and Fig~\ref{fig:adaptive-compression}, by compressing communicated updates, the communication time can be largely reduced. Note that in this paper we only focus on the uplink compression. Because (1) the real-world uplink bandwidth is significantly lower than the downlink bandwidth~\cite{kairouz2021advances,s21124172,NarrowbandIOT}; (2) the FedAvg selects a part of clients instead of all clients in one round, if the aggregated new updates are compressed, those unselected clients cannot receive the newest updates in time. The down-link compression requires a more dedicated design, which is still a challenge in FL compression~\cite{philippenko2020artemis,haddadpour2020federated,babakniya2022federated}. In this paper, we mainly study the adaptive communication compression and weights based on heterogeneous bandwidth.

Fig.~\ref{fig:adaptive-compression} shows that different compression ratios result in different information retained in the compressed parameters. With such adaptive compression based on bandwidth, clients with higher communication bandwidth are assigned with lower compression ratios thus keeping more information to accelerate training convergence, while the communication time of this adaptive compression will be no larger than the uniform compression (shown in Fig.~\ref{fig:Straggler}). 

\section{Method}
\label{Sec:method}
\subsection{System Overview}
\subsubsection{Overview of the system}
\label{Section}
We propose an innovative modification of the FedAvg algorithm, integrating model update sparsification to enhance communication efficiency in FL environments. Given the bandwidth and latency of each client, we set a baseline compression ratio for the slowest selected clients. After local training, the algorithm applies a TOP-K sparsification to the model differences, with compression ratios tailored to equalize communication times across clients. The averaging coefficients of clients are also adjusted according to normalized compression ratios. Notably, the algorithm increases the weight of retained parameters unique to individual clients during the averaging process, ensuring a more representative and efficient global model update. 
We summarize the FedAvg and proposed algorithms in \hyperref[alg:algorithm1]{\textbf{Algorithm 1}}.

\begin{algorithm}[h]
\caption{Summary of FedAvg, Bandwidth-aware Compression Ratio Scheduling (BCRS) and Overlap-aware Parameter Weighted Average (OPWA)}
\label{alg:algorithm1}
\begin{algorithmic}[1]  
\State \textbf{Input:} local datasets $D_i$, number of parties $N$, selected clients $C$, number of communication rounds $T$, number of local epochs $E$, server learning rate $\alpha$, compression ratios $CR$, parameter enlarge rate $\gamma$, learning rate $\eta$
\State \textbf{Output:} The final model $w_T$
\State
\State \textbf{Server executes:}
\State Initialize $w_0$
\For{$t = 0$ to $T-1$}
    \State Sample a set of parties $\mathcal{S}_t$
    \State $n \gets \sum_{i \in \mathcal{S}_t} |D_i|$
    \For{each $i \in \mathcal{S}_t$ \textbf{in parallel}}
        \State Send the global model $w_t$ to party $P_i$
        \State $\Delta w_{it} \gets$ LOCALTRAINING$(i, w_t)$
        \State $\Delta w_{it}^{\text{sparse}} \gets \Call{TopK}{\Delta w_{it}, CR_i}$ 
    \EndFor
    \State $f_i \gets \frac{|D_i|}{n}$
    \State $w_{t+1} \gets w_t - \eta {\sum_{i \in \mathcal{S}_t} f_i \Delta w_{it}^{\text{sparse}}}$ (FedAvg)
    \State $p_i^{\prime}$ =  $\frac{f_i}{\max(f_i, Norm(CR_i)} \times \alpha$ (both BCRS and OPWA)
    \State $w_{t+1} \gets w_t - \eta {\sum_{i \in \mathcal{S}_t} p_i^{\prime} \Delta w_{it}^{\text{sparse}}}$ (BCRS)
    \State calculate mask $M$ (OPWA)
    \State $w_{t+1} \gets w_t - \eta {\sum_{i \in \mathcal{S}_t} p_i^{\prime} \cdot M(\Delta w_{it}^{\text{sparse}})}$ (OPWA)
\EndFor
\State \Return $w_{T}$
\State 
\State \textbf{LOCALTRAINING}$(i, w_t):$
\State $w_{it} \gets w_t$
\For{epoch $k = 1, 2, \ldots, E$}
    \For{each batch $b = \{(x, y)\}$ of $D_i$}
        \State $w_{it} \gets w_{it} - \eta \nabla L(w_{it}; b)$
    \EndFor
\EndFor
\State $\Delta w_{it} \gets w_t - w_{it}$
\State \Return $\Delta w_{it}$ to the server
\vspace{-0.0cm}
\end{algorithmic}
\end{algorithm}

\vspace{-0.0cm}
\subsubsection{Bandwidth-aware Compression Ratio Scheduling (BCRS)} 
\label{Section:BCRS}
Besides inconsistent hardware capabilities, heterogeneous bandwidths also compound the straggler problem since clients with faster transmission speeds are delayed by the need to wait for slower clients before progressing to the next communication round. To address this, we propose the BCRS algorithm that dynamically adjusts the compression ratios in response to bandwidth heterogeneity. By setting the slowest clients’ post-compression communication time as the benchmark, we allocate lower compression ratios to clients with better network capabilities, transmitting more non-zero parameters to update the global model. This approach enables clients to transmit the compressed models at a similar time, mitigating the straggler problem's impact on the model convergence rate. Details are well-explained in \hyperref[alg:algorithm2]{\textbf{Algorithm 2}} and Section~\ref{sec:IV-B}.

\begin{algorithm}
\caption{Bandwidth-aware Compression Ratio Scheduling}
\label{alg:algorithm2}
\begin{algorithmic}[1]  
\State \textbf{Input:} number of clients $N$, selected clients $\mathcal{S}_t$, model update size $V$, $i$-th client's bandwidth $B_i$, $i$-th client's latency $L_i$, default compression ratio $CR^{*}$
\State \textbf{Output:} List containing Compression Ratios $CR$
\State \textbf{CalculateCR}$(CR^{*}):$
\State Initialize $CR \gets$ empty list
\State Initialize $idx_{max},T_{max} \gets 0$
\For{each client $i \in \mathcal{S}_t$}
    \State $T_{comm,i} \gets L_i + \frac{2 \times V \times CR*}{B_i}$
    \If{$T_{comm,i} > T_{max}$}
        \State $T_{max} \gets T_{comm,i}$
        \State $idx_{max} \gets i$
    \EndIf
\EndFor
\State $T_{bench} \gets T_{max}$
\For{each client $i \in \mathcal{S}_t$}
    \State $CR_i \gets \left( \frac{T_{bench} - L_i}{2 \times V} \right) \times B_i$
    \State Append $CR_i$ to $CR$
\EndFor
\State \Return $CR$
\end{algorithmic}
\end{algorithm}

\vspace{-0.2cm}
\begin{figure}[ht]
\setlength{\abovedisplayskip}{-1pt}
    \subfigbottomskip=2pt
    \subfigcapskip=1pt
    \setlength{\abovecaptionskip}{0.cm}
\vspace{-0.20cm}
\centering
	\includegraphics[width=1.00\linewidth]{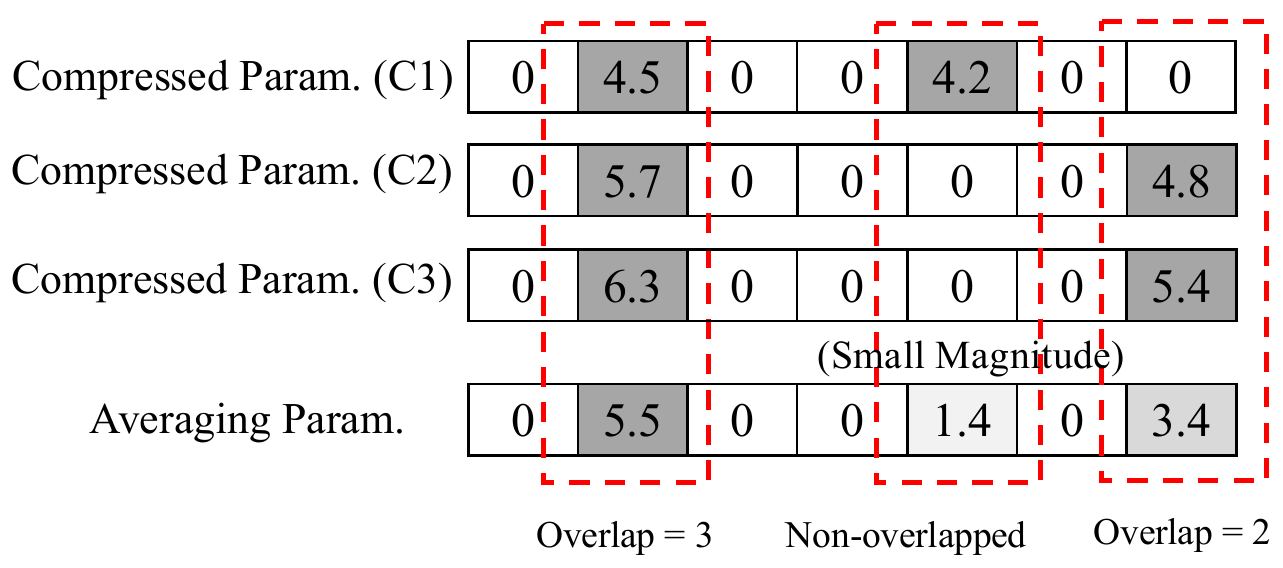}
	\caption{Illustration of Parameter Overlap: Smaller magnitude of less overlapped parameters compared to overlapped parameters after averaging. Param represents the model parameters,  C1, C2, and C3 represent different clients.}
	\label{fig:overlapless}
\vspace{-0.5cm}
\end{figure}
\subsubsection{Overlap-aware Parameter Weighted Average (OPWA)}
\label{Section:OPWA}
Upon examining the retained parameters after compression, a notable observation emerges: the parameter retention patterns across clients exhibit significant heterogeneity, with a substantial number of parameters appearing only occasionally, as illustrated in Fig~\ref{fig:distribution_pattern}. Motivated by this observation, we define a new metric \textit{Degree of Overlap}, which quantifies the frequency of a specific parameter's presence in the compressed model updates of selected clients. This metric is well-explained in Fig.~\ref{fig:overlapless}. To address the pattern heterogeneity, we propose OPWA, adding a parameter mask that amplifies the weights of parameters with a lower degree of overlap based on BCRS. Details are explained in \hyperref[alg:algorithm3]{\textbf{Algorithm 3}} and Section~\ref{sec:IV-C}.

\subsection{BCRS}
\label{sec:IV-B}

Magnitude pruning has been a prevalent compression method used in compressed DNNs, where model parameters with small magnitudes are eliminated. As mentioned above, this method still suffers from the straggler problem. To mitigate these issues, we propose an adaptive bandwidth-aware compression method to fully utilize the waiting time of faster clients to transmit more parameters by assigning a lower compression ratio. 

Each client is initialized with a specific bandwidth and latency, and the communication time with the uniform compression ratio is calculated using our calculation method, mentioned in Section~\ref{Section:FL}. The slowest client's post-compression time is used as a benchmark, calculated by: 
\vspace{-0.1cm}
\begin{equation}
    T_{bench} = \text{argmin}_{i} \left(L + 2 \times \frac{ V_i \times CR_i}{B} \right)
\end{equation}
where $i$ represents the selected clients' indexes.
To align with this benchmark, the compression ratios of other clients are adjusted to utilize their respective bandwidths fully. Furthermore, we calculate an adjusted averaging coefficient with a maximum value of 1: 
\begin{equation}
    p_k^{\prime} =  \frac{f_i}{\max(f_i, Norm(CR_i))} \times \alpha
\end{equation}
where $\alpha$ is the predefined server learning rate.

\subsection{Overlap-aware Parameter Weighted Average}
\label{sec:IV-C}
\begin{figure*}[!h]
\setlength{\abovedisplayskip}{-1pt}
    \subfigbottomskip=2pt
    \subfigcapskip=1pt
    \setlength{\abovecaptionskip}{0.cm}
\vspace{-0.5cm}
\centering
	\subfigure[$\beta$ = 0.1, CR = 0.01]
	{
	\includegraphics[width=0.23\linewidth]{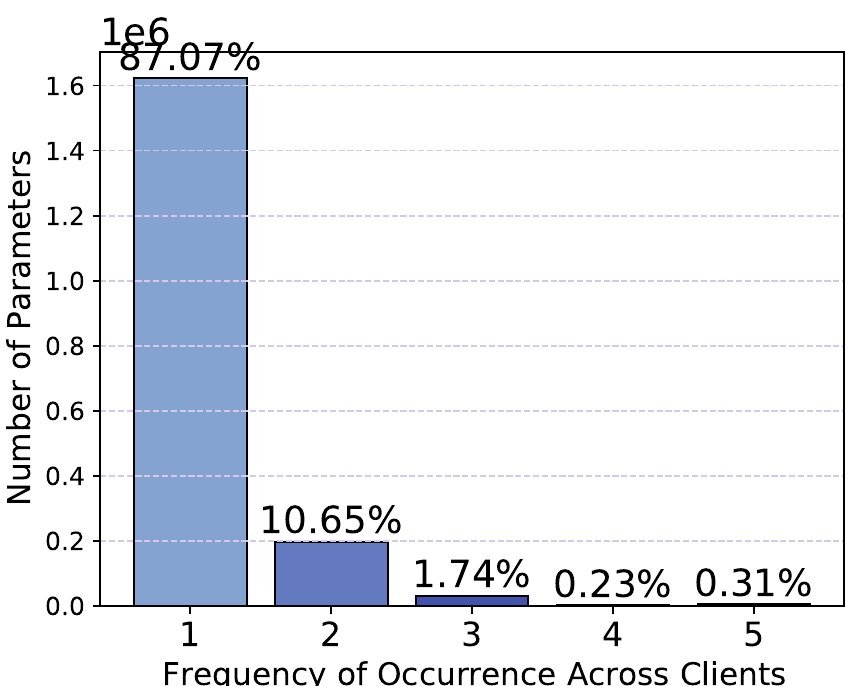}
	}
	\subfigure[$\beta$ = 0.1, CR = 0.1]
	{
	\includegraphics[width=0.23\linewidth]{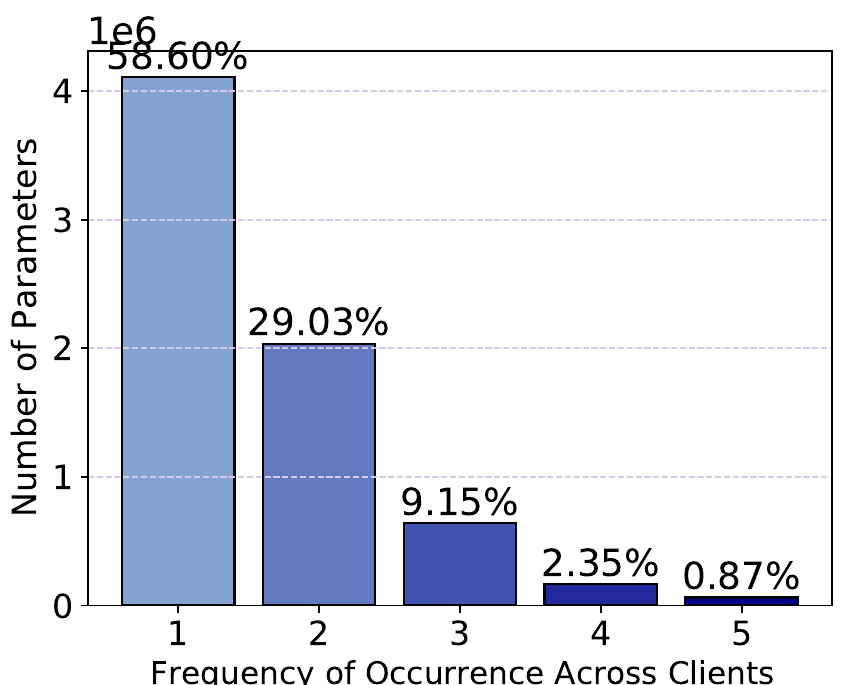}
	}
 \subfigure[$\beta$ = 0.5, CR = 0.01]
	{
	\includegraphics[width=0.23\linewidth]{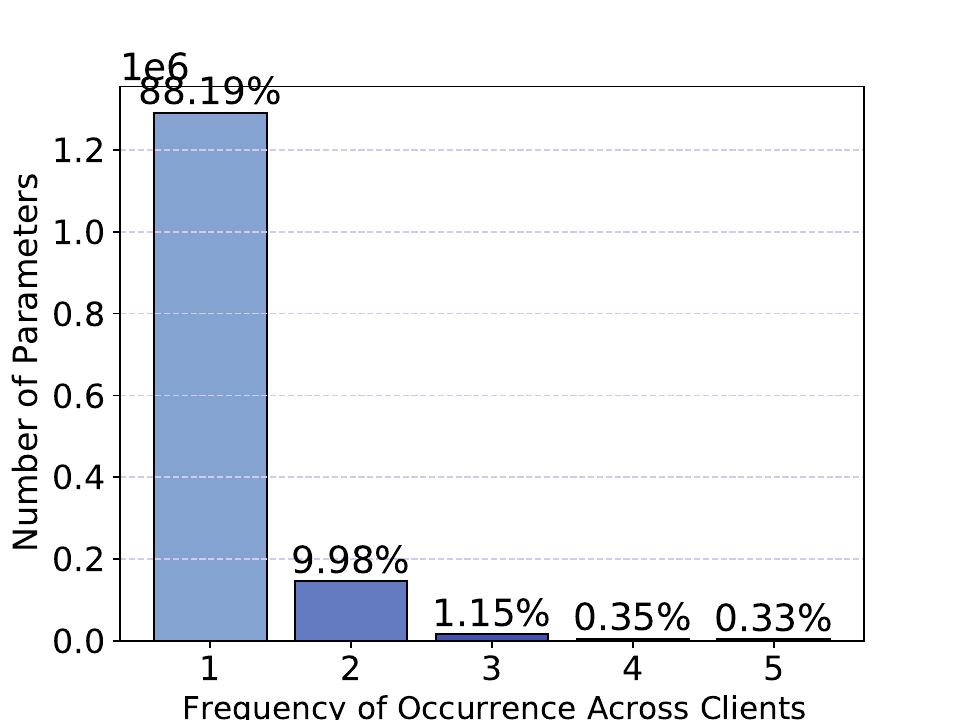}
	}
 \subfigure[$\beta$ = 0.5, CR = 0.1]
	{
	\includegraphics[width=0.23\linewidth]{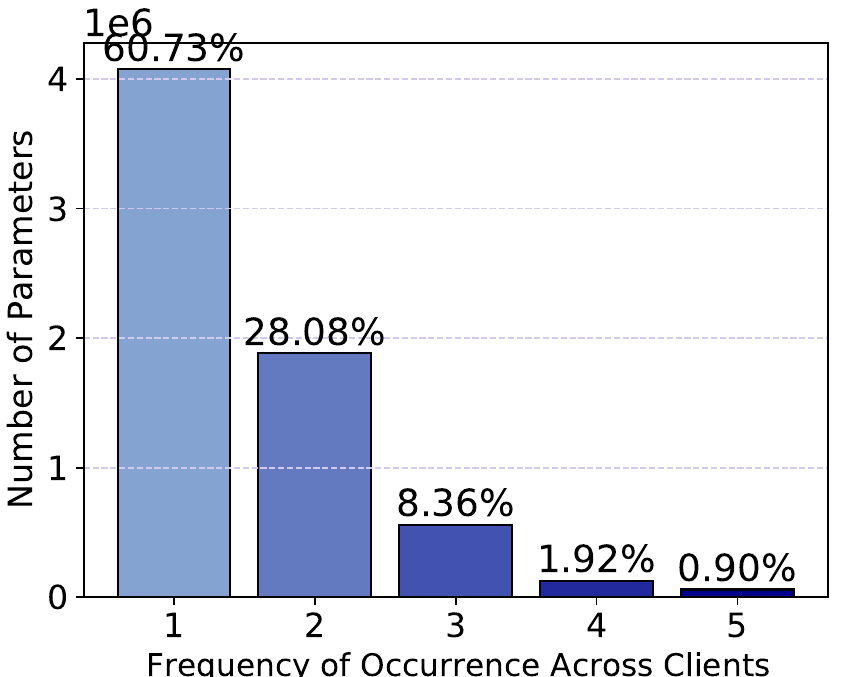}
	}
	\caption{Distribution of degree of overlap of retained parameters after compression.}
	\label{fig:distribution_pattern}
\vspace{-0.3cm}
\end{figure*}

As illustrated in Fig~\ref{fig:distribution_pattern}, we observe that severe parameter retention pattern heterogeneity happens when the model updates are compressed with high compression ratios. 
Half of the retained parameters appear only once in the compressed model updates of selected clients under CR = 0.1, leading to under-updating in the conventional averaging process. This phenomenon is exacerbated in a high compression level (CR = 0.01). This situation conceptually mirrors a low learning rate on the server for parameters with minimal overlap, which are crucial in reflecting the unique characteristics of non-IID local datasets.

Inspired by the retention pattern heterogeneity, we modified the averaging process of BCRS by adding a parameter-wise mask $M$ that adjusts the weights for parameters with low degrees of overlap, thereby balancing their contribution against those that are more frequently updated: 
\begin{equation}
    w_{t+1} \gets w_t - \eta {\sum_{i \in S_t} p_i^{\prime} \cdot M(\Delta w_{it}^{\text{sparse}})}
\end{equation}
Implementation details can be checked in \hyperref[alg:algorithm3]{\textbf{Algorithm 3}}.

\begin{algorithm}
\label{alg:algorithm3}
\caption{Overlap-aware Parameter Weighted Average (OPWA)}
\begin{algorithmic}[1]
\State \textbf{Input:} selected clients $\mathcal{S}_t$, $i$-th client's compressed model updates $w_{i}^{\text{sparse}}$, enlarge rate $\gamma$, required degree of overlap $D$ (set to be 1 by default)
\State \textbf{Output:} Mask $M$
\State
\State \textbf{CalculateOverlap:}
\State Initialize $overlapdict$ as an empty dictionary
\For{$i = 0$ to $\left| \mathcal{S}_t\right|-1$}
    \State $w_{i,flatten}^{\text{sparse}} \gets flatten \left( w_{i}^{\text{sparse}} \right)$
    \For{param $p$ in $w_{i,flatten}^{\text{sparse}}$}
        \If{$p$ $exists$}
            \State $overlapdict[p] \gets overlapdict[p] + 1$
        \EndIf
    \EndFor
\EndFor
\State \Return $overlapdict$
\State 
\State \textbf{GenerateMask($overlapdict$):}
\State Initialize $M$ as an empty dictionary
\For{param $p$ in $overlapdict$}
    \If{$overlapdict[p] \leq D$}
        \State $M[p] \gets \gamma$
    \Else 
        \State $M[p] \gets 1$
    \EndIf
\EndFor
\State \Return $M$
\end{algorithmic}
\end{algorithm}
\vspace{-0.2cm}

\section{Experiment}
\label{Sec:Experiment}
\subsection{Experiment Setup}
\label{Sec:params}
\textbf{Federated Datasets and Models.} To evaluate the effectiveness of our bandwidth-aware compression algorithm with overlap-weighted averaging, we conduct extensive experiments on three commonly used datasets~\cite{caldas2018leaf,longtailinfocom,VHL}: CIFAR-10, CIFAR-100, and SVHN. These datasets are evaluated using the ResNet18 model~\cite{resnet}. 

\begin{figure}[H]
\setlength{\abovedisplayskip}{-1pt}
    \subfigbottomskip=2pt
    \subfigcapskip=1pt
    \setlength{\abovecaptionskip}{0.1cm}
\vspace{-0.5cm}
\centering
	\subfigure[$\beta$ = 0.5]
	{
	\includegraphics[width=0.46\linewidth]{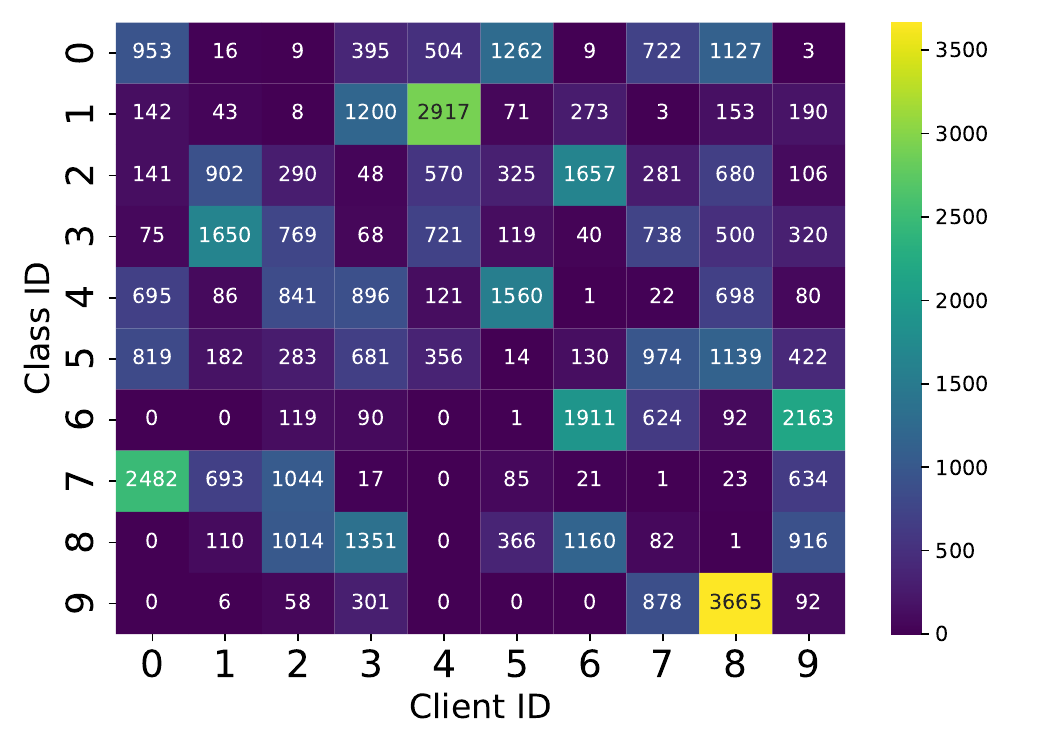}
	}
	\subfigure[$\beta$ = 0.1]
	{
	\includegraphics[width=0.46\linewidth]{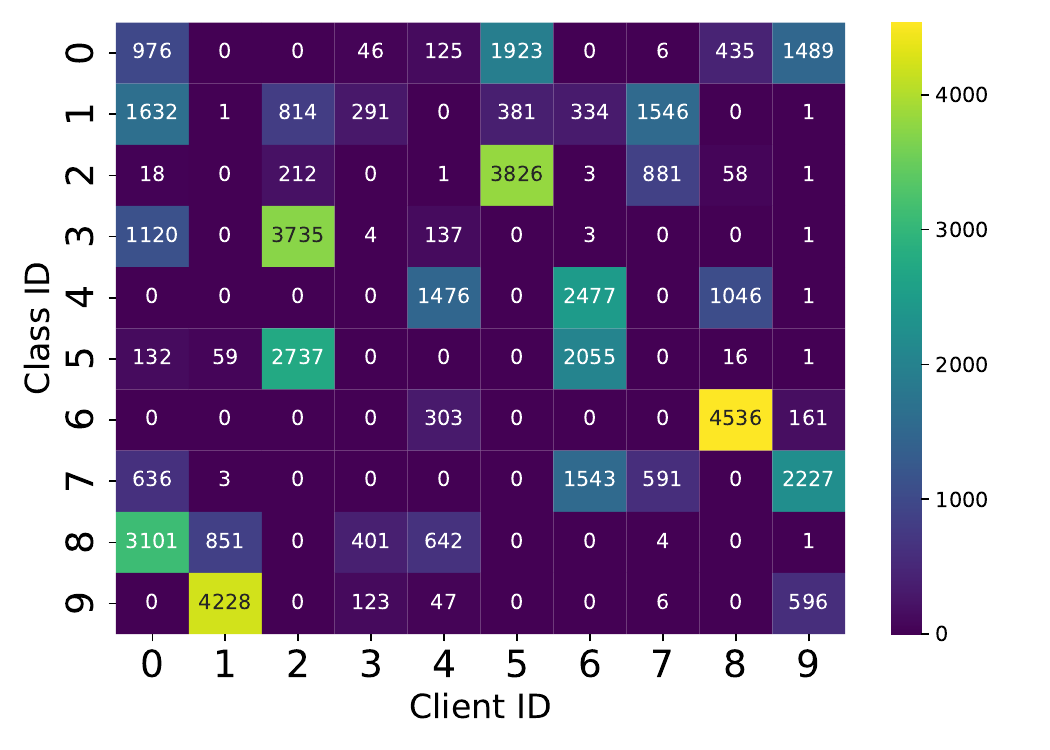}
	}
	\caption{NIID Distribution Across Clients for CIFAR-10.}
	\label{fig:datahetero}
\vspace{-0.4cm}
\end{figure}

\textbf{Federated settings.} We adopt the settings used in \cite{li2021federatedstudy, VHL} that use distribution-based label-skew partitioning to simulate non-IID conditions. We allocate each client a proportion of data samples of each label following the Dirichlet distribution, a commonly used prior distribution in Bayesian inference. The distribution follows $p_k \sim Dir(\beta)$ and $i$-th client is allocated a $p_{k, i}$ proportion of data samples of class $k$. In our experiments, we use $\beta = 0.1,0.5$ to simulate severe and moderate data heterogeneity, as shown in Fig~\ref{fig:datahetero}.
Experiments are conducted under $N = 10$, client participation ratio $C = 0.5$, batch size $bs = 64$, communication rounds = 200, and epochs $E = 1$. We also use $N=16, 20$ to test the scalability. All experiments are conducted using NVIDIA GeForce RTX 4090s.

\textbf{Baselines.} We compare BCRS and BCRS+OPWA algorithms with FedAvg and its sparsified variants, including TOPK and Error-feedback TOPK \cite{haddadpour2021federated,sattler2019robust,pmlr-v202-li23o}.
\vspace{-0.2cm}
\subsection{Measurements}
\label{Section5B:Measurement_Bandwidth_latency}
\textbf{Bandwidth and Latency.} 
Clients are initialized with randomly generated bandwidth with a mean of 1 Mbits/s and a standard deviation of 0.2 Mbits/s in a normal distribution. The latencies of clients are uniformly distributed with a range of (50ms, 200ms].

\textbf{Time Metric and Comparision.} 
We define and accumulate three metrics over total communication rounds for evaluating communication efficiency.
\begin{itemize}
        \item \textit{Actual Time:} The actual communication time in a round. The accumulation reflects model transmission durations.
        \item \textit{Maximum Communication Time:} The actual communication duration due to the straggler. The accumulation represents the total transmission duration of FedAvg.
        \item \textit{Minimum Communication Time:} Indicative of the fastest client's communication time without straggler. The accumulation indicates the optimal scenario.
    \end{itemize}

\subsection{Experiment Results}

\subsubsection{Main Result}
We have conducted extensive experiments under different compression levels: $CR = 0.1$ and $CR = 0.01$ and degrees of data heterogeneity: $\beta = 0.1$ and $\beta = 0.5$. 
The final model's test accuracy of our proposed BCRS and OPWA algorithms, alongside the baselines FedAvg, TOPK, and EFTOPK, across various datasets, are summarized in Table~\ref{tab:my_table}.

\begin{table*}[!h]
    \centering
    \caption{Main Results of Algorithms on different datasets: Test accuracies under the same setting in Section~\ref{Sec:params}.}
    \label{tab:my_table}
    \begin{tabular}{|c|c|c|c|c|c|c|c|c|c|c|c|c|}
        \hline\hline
        Datasets & \multicolumn{4}{c}{CIFAR-10} & \multicolumn{4}{c}{SVHN} & \multicolumn{4}{c}{CIFAR-100} \\ \hline\hline
        Data Heterogeneity & \multicolumn{2}{|c|}{$\beta=0.1$} & \multicolumn{2}{|c|}{$\beta=0.5$} & \multicolumn{2}{|c|}{$\beta=0.1$} & \multicolumn{2}{|c|}{$\beta=0.5$} & \multicolumn{2}{|c|}{$\beta=0.1$} & \multicolumn{2}{|c|}{$\beta=0.5$} \\ \hline
        Compression Ratio & 0.1 & 0.01 & 0.1 & 0.01 & 0.1 & 0.01 & 0.1 & 0.01 & 0.1 & 0.01 & 0.1 & 0.01\\ \hline
        FedAvg (Uncompressed)&  0.568&  0.568&  0.7637&  0.7637&  0.6235&  0.6235&  0.9113&  0.9113 & 0.4921 & 0.4921 & 0.5686 & 0.5686\\ \hline
        TOPK \cite{sattler2019robust,haddadpour2021federated}&  0.4669&  0.2555&  0.6853&  0.3268&  0.4052&  0.304&  0.8905&  0.7771 & 0.4234 & 0.2418 & 0.4965 & 0.2616\\ \hline
        EFTOPK \cite{pmlr-v202-li23o} &  0.4553&  0.247&  0.6848&  0.3123&  0.5151&  0.264&  0.8918&  0.7738 & 0.4262 & 0.2504 & 0.4962 & 0.2629\\ \hline
        BCRS (ours)&  0.493&  0.305&  0.7124&  0.4828&  0.6619&  0.3493 &  0.8925&  0.7945 & 0.2382 & 0.3053 & 0.5415 & 0.4345\\ \hline
        BCRS+OPWA (ours)&  \textbf{0.6029}&  \textbf{0.4845}&  \textbf{0.7437}&  \textbf{0.5528}&  \textbf{0.7063}&  \textbf{0.5259}&  \textbf{0.9031}&  \textbf{0.8728} & \textbf{0.4892} & \textbf{0.4775} & \textbf{0.5499} & \textbf{0.4966}\\ \hline
    \end{tabular}
\end{table*}

\subsubsection{Evaluation of Bandwidth-Aware Compression Ratio Scheduling (BCRS)} 
To assess the effectiveness of the BCRS algorithm, we test the algorithm with CR = (0.1, 0.01) under $\beta$ = (0.1, 0.5) and compared these results against TOPK and EFTOPK and the uncompressed FedAvg under identical FL settings. It's worth noting that in all the experiments, the hyperparameter $\alpha$ of the BCRS algorithm is tuned across a set of candidate values \{0.01, 0.03, 0.1, 0.3, 1\} to identify the optimal configuration for each scenario.

\begin{figure}[!htb]
\setlength{\abovedisplayskip}{-1pt}
    \subfigbottomskip=2pt
    \subfigcapskip=1pt
    \setlength{\abovecaptionskip}{0.2cm}
\vspace{-0.5cm}
\centering
	\subfigure[CR=0.01]
	{
	\includegraphics[width=0.47\linewidth]{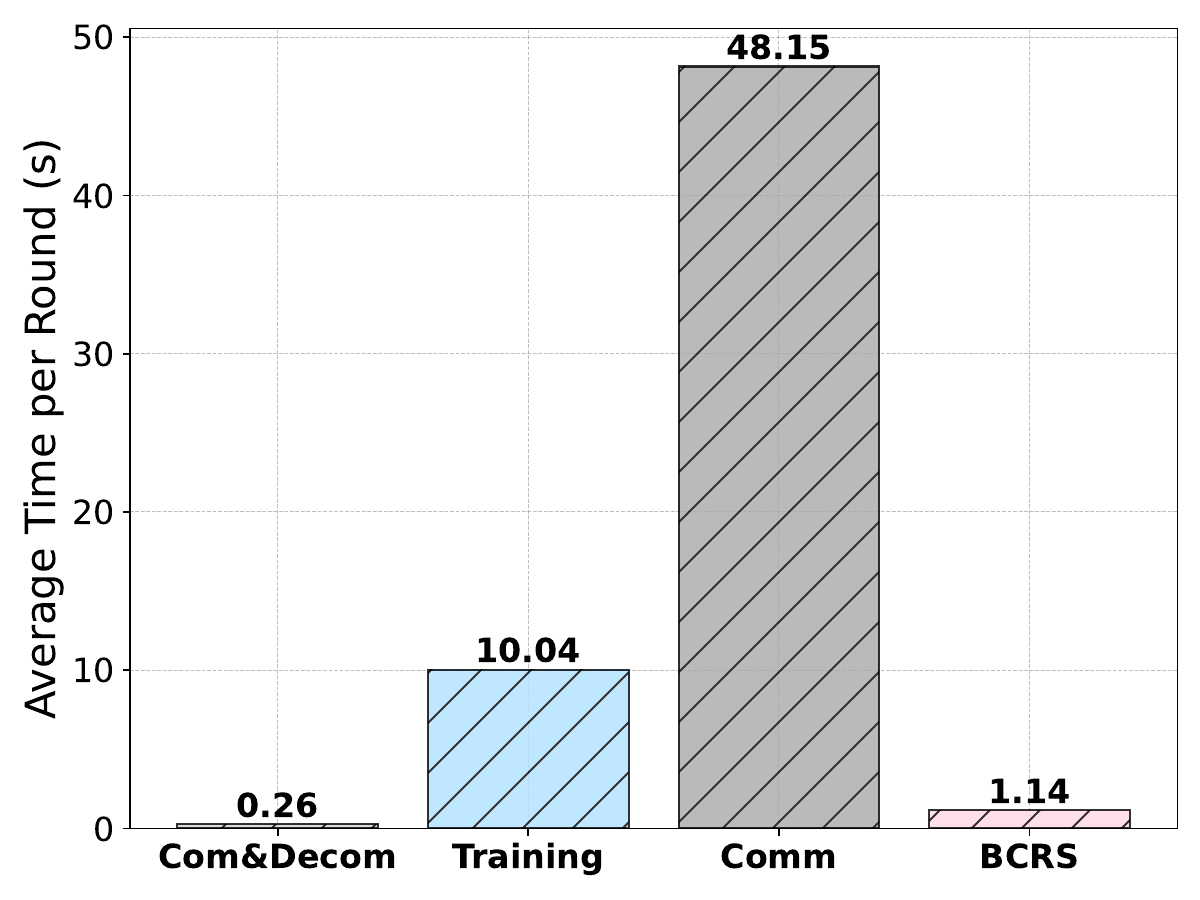}
	}
	\subfigure[CR=0.1]
	{
	\includegraphics[width=0.47\linewidth]{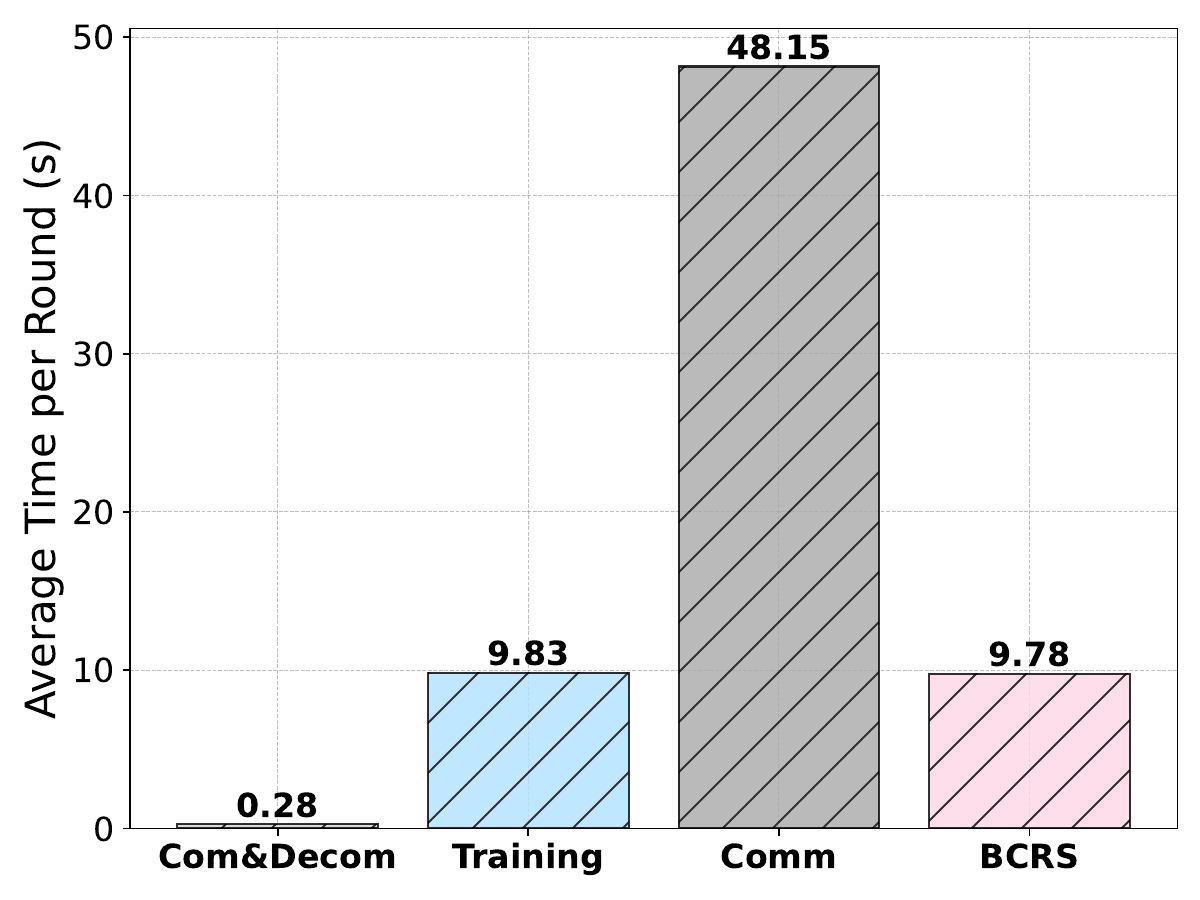}
	}
	\caption{Time Breakdown in one FL round: Compress\& Decompress, Training, Uncompressed Communication, BCRS Communication. }
	\label{fig:breakdown}
\vspace{-0.5cm}
\end{figure}

\begin{figure}[!htb]
\setlength{\abovedisplayskip}{-1pt}
    \subfigbottomskip=0pt
    \subfigcapskip=0pt
    \setlength{\abovecaptionskip}{0.cm}
\vspace{-0.5cm}
\centering
	\subfigure[$\beta$ = 0.1, CR = 0.1]
	{
	\includegraphics[width=0.46\linewidth]{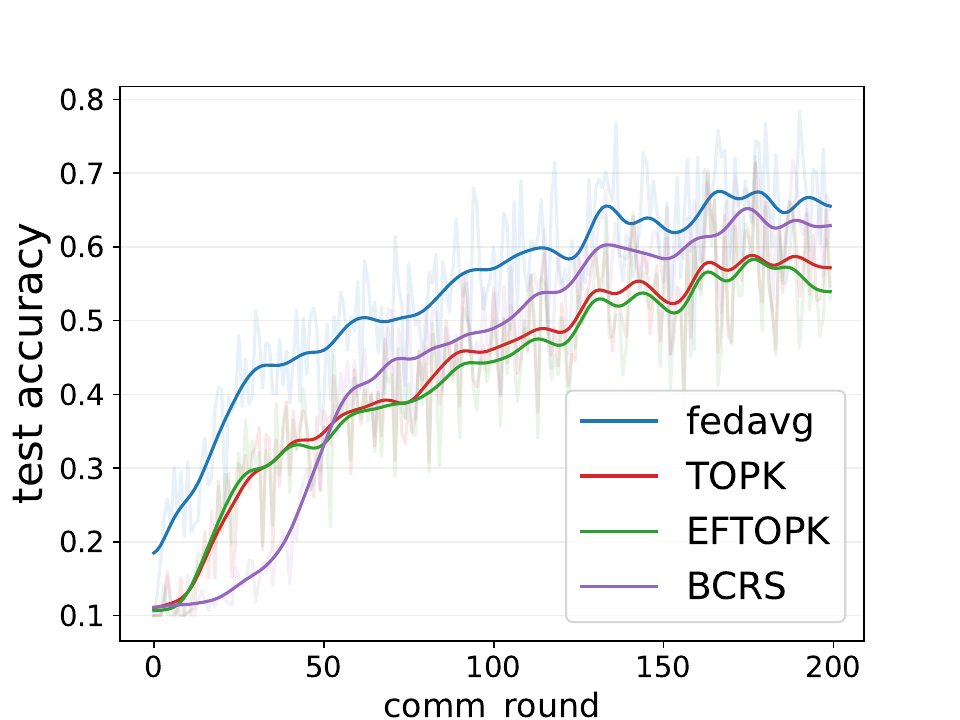}
	}
	\subfigure[$\beta$ = 0.5, CR = 0.1]
	{
	\includegraphics[width=0.46\linewidth]{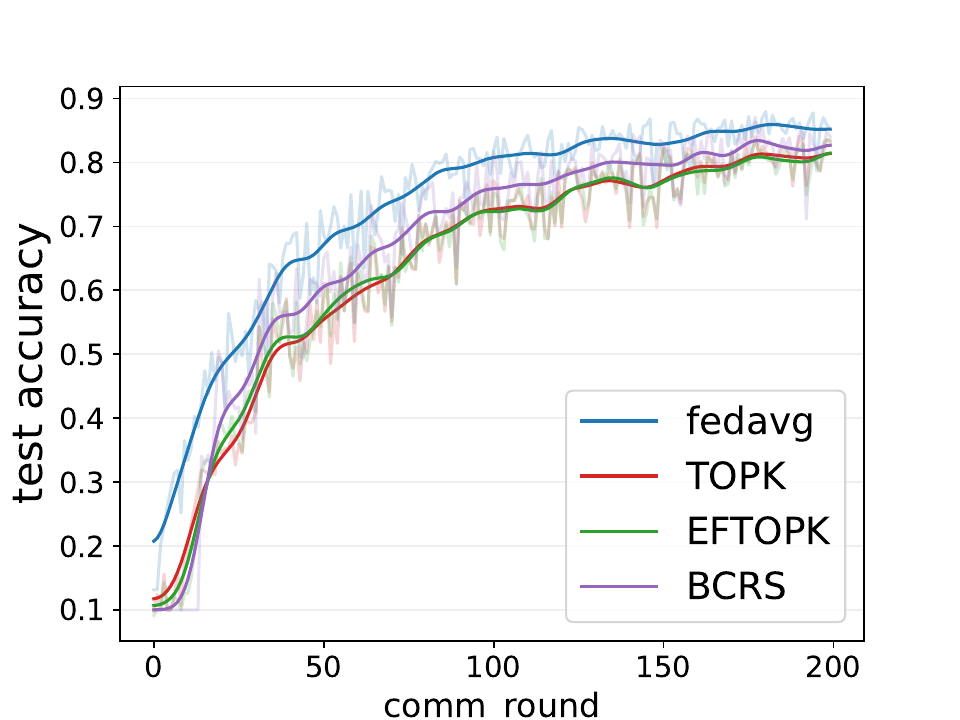}
	}
 \subfigure[$\beta$ = 0.1, CR = 0.01]
	{
	\includegraphics[width=0.46\linewidth]{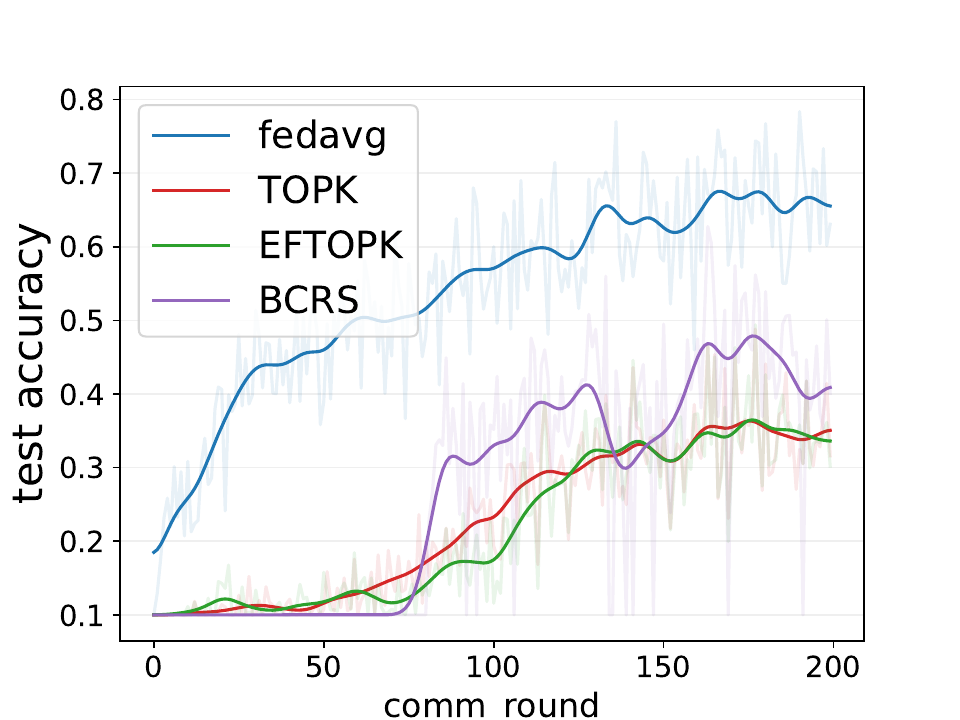}
	}
 \subfigure[$\beta$ = 0.5, CR = 0.01]
	{
	\includegraphics[width=0.46\linewidth]{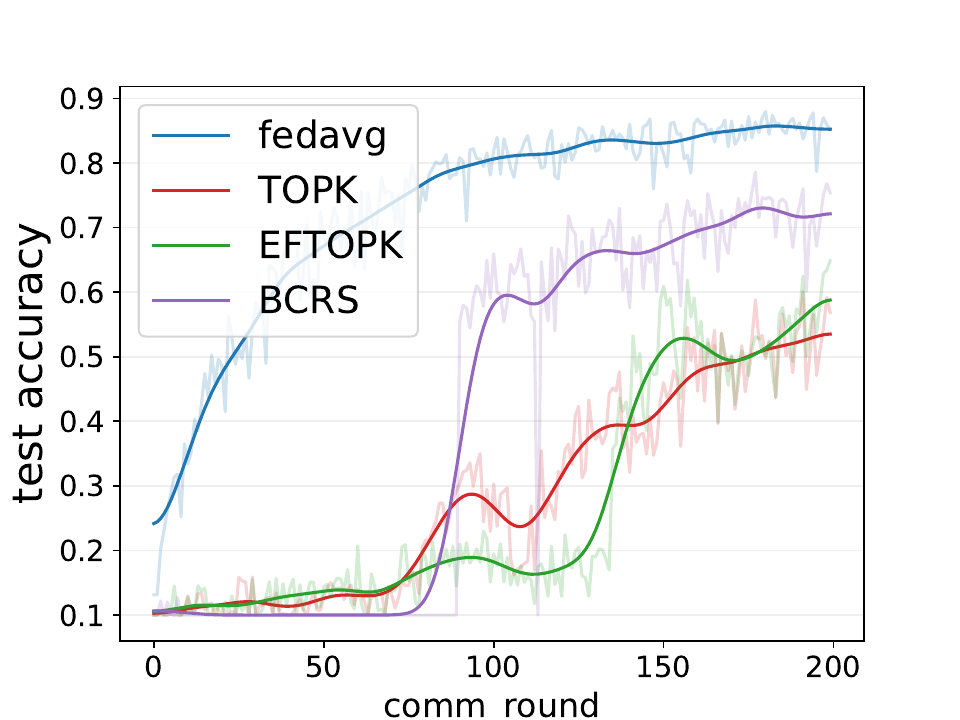}
	}
	\caption{CIFAR-10: BCRS with other baselines.}
	\label{fig:overlap_best_01}
\vspace{-0.0cm}
\end{figure}

\begin{figure}[!h]
\setlength{\abovedisplayskip}{-1pt}
    \subfigbottomskip=0pt
    \subfigcapskip=0pt
    \setlength{\abovecaptionskip}{0.1cm}
\vspace{-0.5cm}
\centering
	\subfigure[$\beta$ = 0.1, CR = 0.1]
	{
	\includegraphics[width=0.46\linewidth]{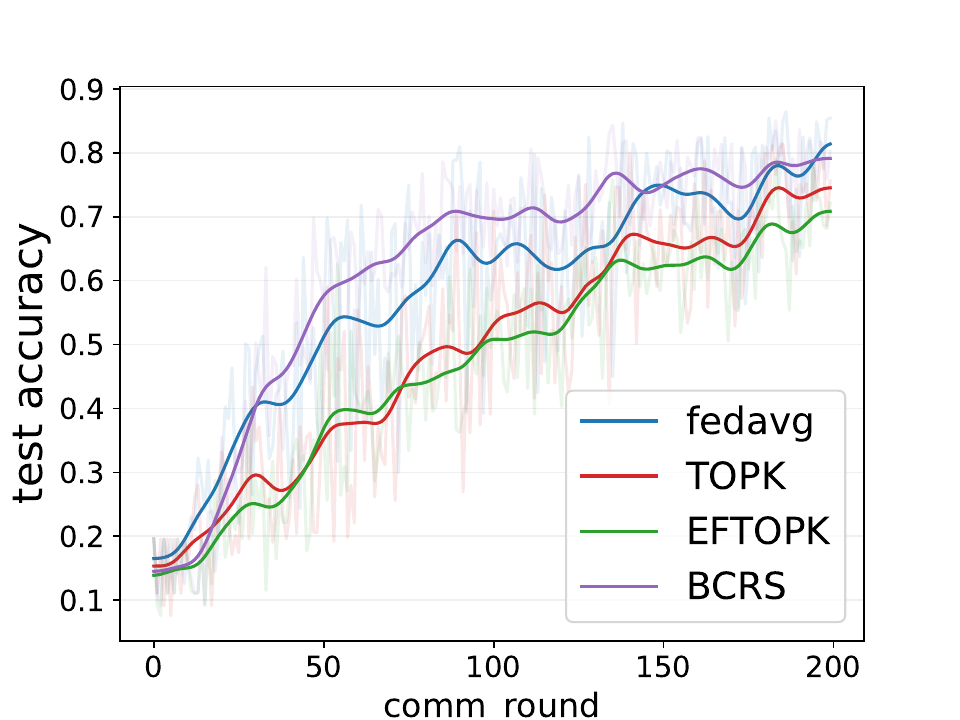}
	}
	\subfigure[$\beta$ = 0.1, CR = 0.01]
	{
	\includegraphics[width=0.46\linewidth]{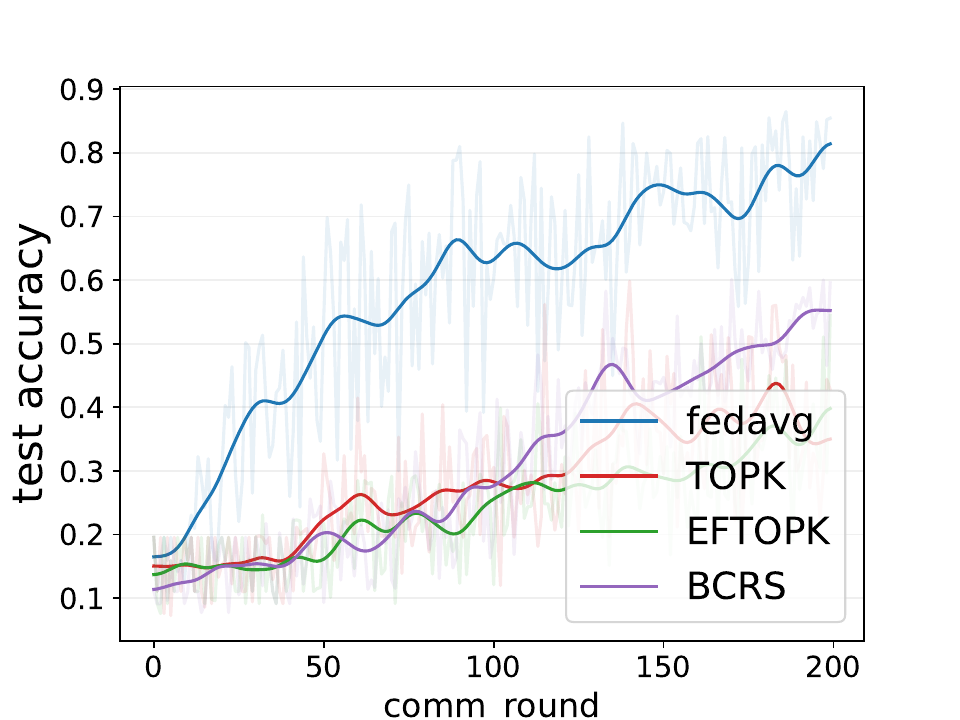}
	}
 \subfigure[$\beta$ = 0.5, CR= 0.1]
	{
	\includegraphics[width=0.46\linewidth]{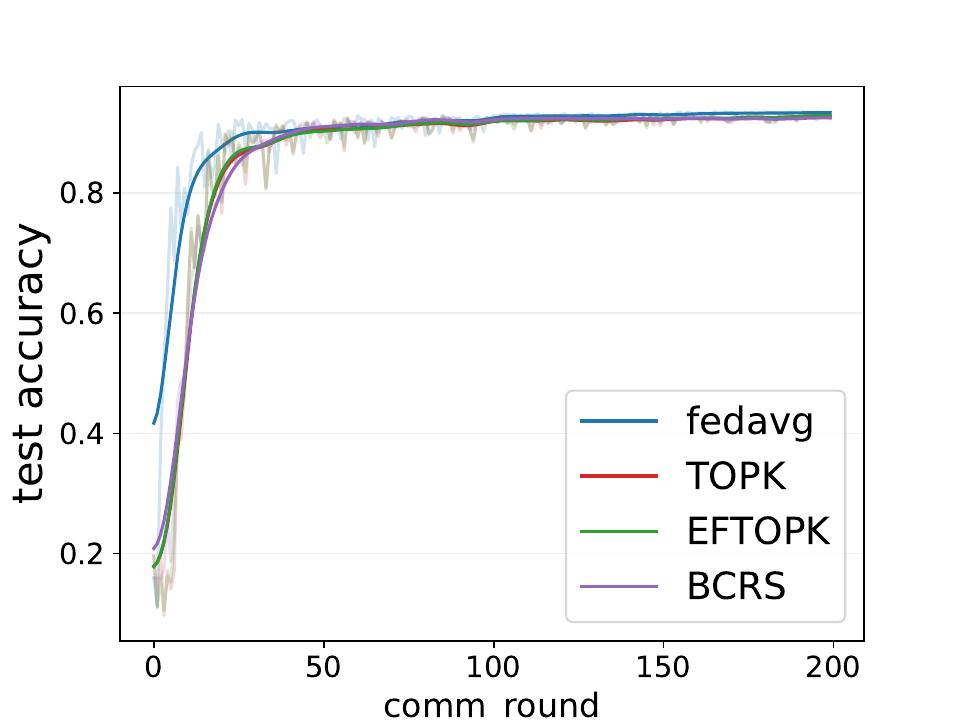}
	}
 \subfigure[$\beta$ = 0.5, CR = 0.01]
	{
	\includegraphics[width=0.46\linewidth]{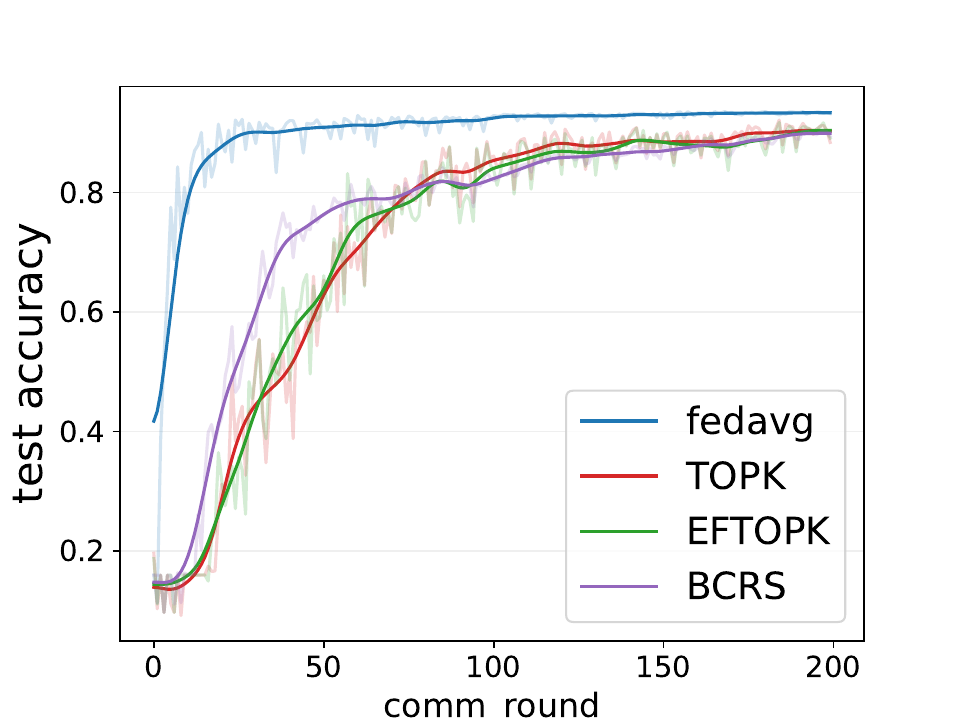}
	}
	\caption{SVHN: Comparison between BCRS and Baselines.}
	\label{fig:svhnB}
\vspace{-0.3cm}
\end{figure}

\begin{figure}[!h]
\setlength{\abovedisplayskip}{-1pt}
    \subfigbottomskip=0pt
    \subfigcapskip=0pt
    \setlength{\abovecaptionskip}{0.1cm}
\vspace{-0.45cm}
\centering
\subfigure[$\beta$ = 0.1, CR = 0.1]
	{
	\includegraphics[width=0.46\linewidth]{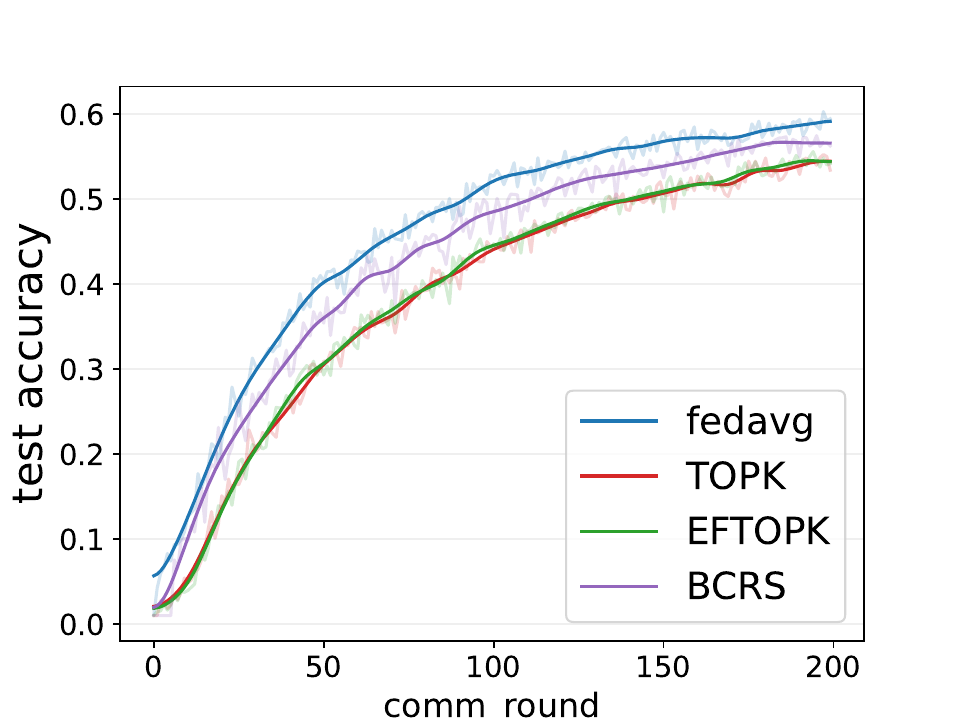}
}
	\subfigure[$\beta$ = 0.1, CR = 0.01]
	{
	\includegraphics[width=0.46\linewidth]{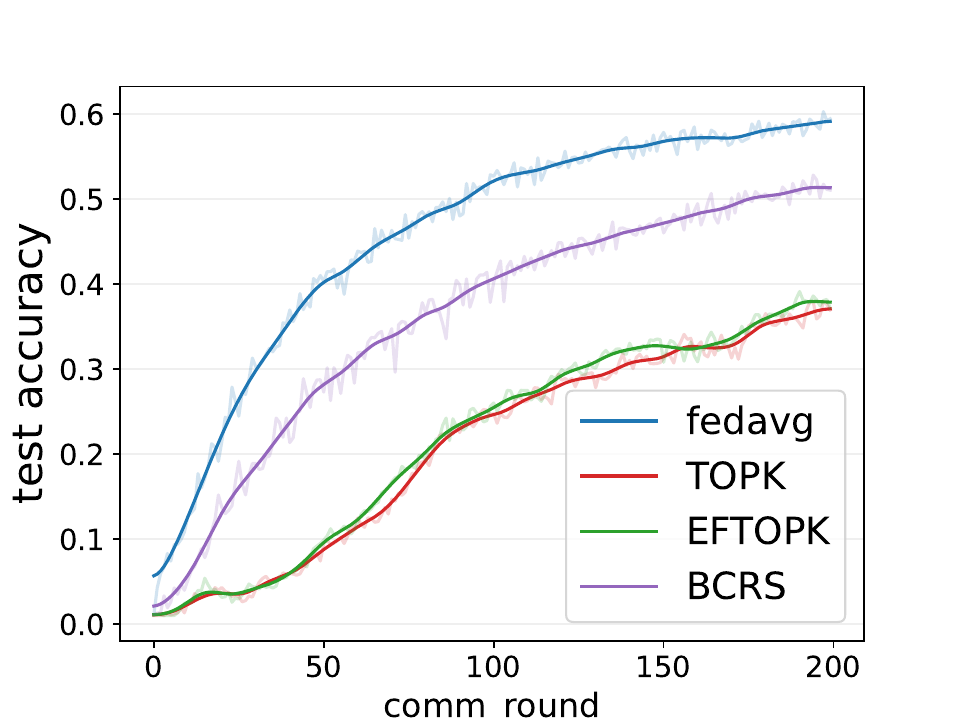}
	}
 \subfigure[$\beta$ = 0.5, CR= 0.1]
	{
	\includegraphics[width=0.46\linewidth]{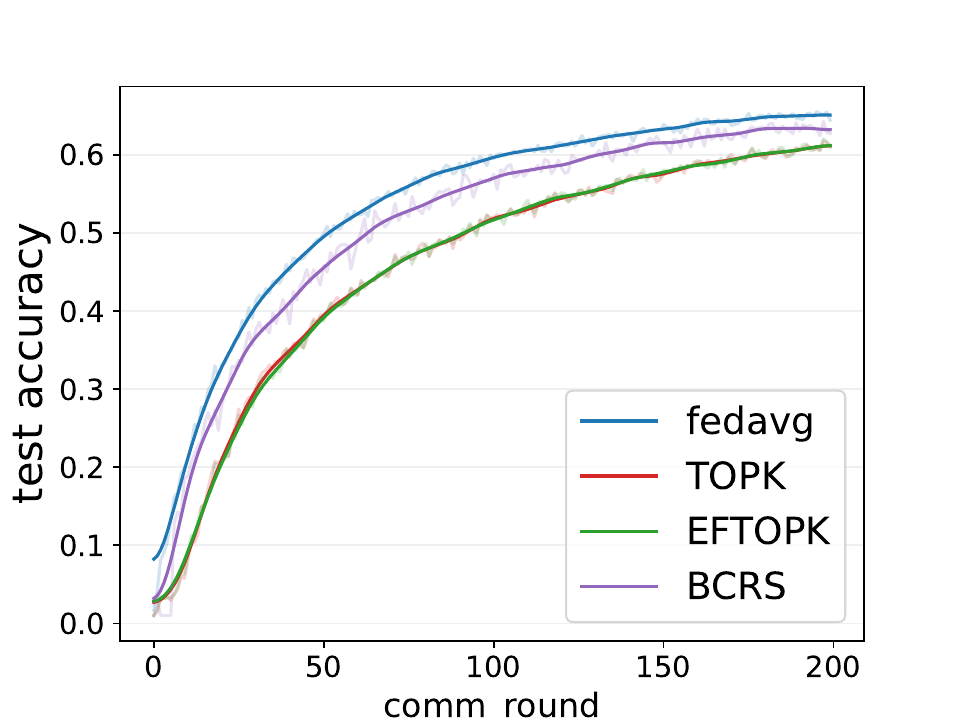}
	}
 \subfigure[$\beta$ = 0.5, CR = 0.01]
	{
	\includegraphics[width=0.46\linewidth]{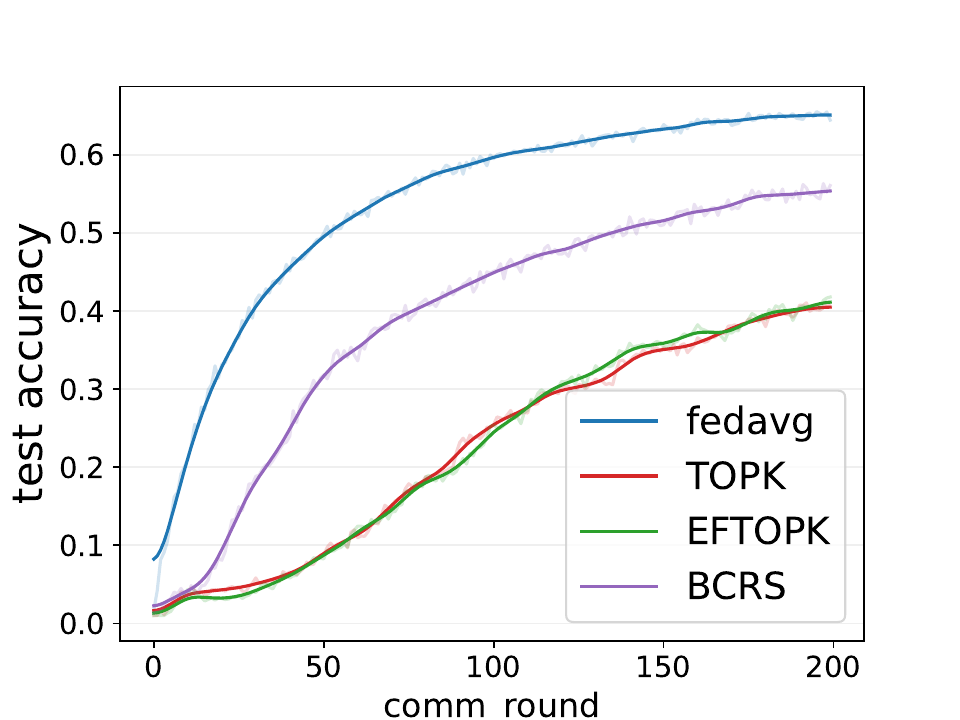}
	}
	\caption{CIFAR-100: BCRS with other baselines.}
	\label{fig:cifar100B}
\vspace{-0.5cm}
\end{figure}

\textbf{Performance of BCRS.}
The above two figures in Fig.~\ref{fig:overlap_best_01} present the results of different heterogeneity settings under $CR = 0.1$. The best server learning rates $\alpha$ are 0.1 and 0.3 for the two settings respectively. The plot demonstrates that the BCRS algorithm maintains a faster convergence rate than the baseline methods. The BCRS algorithm also outperforms other baselines with the optimal server learning rate $\alpha = 0.3$ in both cases. Results on SVHN and CIFAR-100 are shown in Fig.~\ref{fig:svhnB} and Fig.~\ref{fig:cifar100B}.

\textbf{Communication Efficiency.} In evaluating the communication efficiency of both baselines and BCRS, we record the time to reach target accuracy for each algorithm on CIFAR-10 as illustrated in Table~\ref{table:communication_efficiency}. It can be observed that there's a huge difference between the accumulated $Min Time$ and $Max Time$, highlighting the need to mitigate the straggler problem. To achieve 40\% accuracy, FedAvg, as a standard baseline, required approximately 3677.238 seconds to reach this accuracy level. Under $CR=0.1$, the TOPK and EFTOPK algorithms demonstrated comparable efficiency, taking about 281.364 seconds and 157.412 seconds, respectively. In stark contrast, our BCRS algorithm significantly improves performance efficiency, achieving the target accuracy in just 17.938 seconds, much faster than the baselines. 
Fig.~\ref{fig:breakdown} displays the breakdown of each process in one FL round and our BCRS algorithm effectively mitigates the communication overhead.
To clarify, blanks in Table~\ref{table:communication_efficiency} do not mean missing experiments but indicate instances that are not applicable to the experiments. Blanks in the $Max Time$ and $Min Time$ columns for BCRS indicate non-applicability since the BCRS algorithm is designed to equalize client communication time. Therefore, specific maximum and minimum time measurements are not meaningful and thus omitted from Table~\ref{table:communication_efficiency}.

\vspace{-0.3cm}
\begin{table}[!ht]
    \centering
    \caption{Communication time (second) to reach the target accuracy (40\%) on CIFAR-10 under $\beta=0.1$. Blanks carry no meaning in the context of our experiments.}
    \begin{tabular}{|p{1.0cm}|c|c|c|c|}
    \hline
        \multicolumn{2}{|c|}{\multirow{2}{*}{Algorithm}} & \multicolumn{3}{c|}{CIFAR-10(40\%)} \\ \cline{3-5}
        \multicolumn{2}{|c|}{} & Actual Time & Max Time & Min Time \\ \hline
        \multirow{2}{*}{FedAvg} & CR=0.1 & 3677.238 & 3677.238 & 104.514 \\ \cline{2-5}
        & CR=0.01 & 3677.238 & 3677.238 & 104.514 \\ \hline
        \multirow{2}{*}{TOPK} & CR=0.1 & 281.364 & 1386.653 & 28.317 \\ \cline{2-5}
        & CR=0.01 & 86.985 & 3634.929 & 74.482 \\ \hline
        \multirow{2}{*}{EFTOPK} & CR=0.1 & 157.412 & 1521.802 & 31.073 \\ \cline{2-5}
        & CR=0.01 & 52.062 & 3719.547 & 76.245 \\ \hline
        \multirow{2}{*}{BCRS} & CR=0.1 & 17.938 & -- & -- \\ \cline{2-5}
        & CR=0.01 & 25.755 & -- & -- \\ \hline
    \end{tabular}
    \label{table:communication_efficiency}
\vspace{-0.4cm}
\end{table}

\begin{figure}[!h]
\setlength{\abovedisplayskip}{-1pt}
    \subfigbottomskip=0pt
    \subfigcapskip=1pt
    \setlength{\abovecaptionskip}{0.1cm}
\vspace{-0.5cm}
\centering
	\subfigure[$\beta$ = 0.1, CR = 0.1]
	{
	\includegraphics[width=0.46\linewidth]{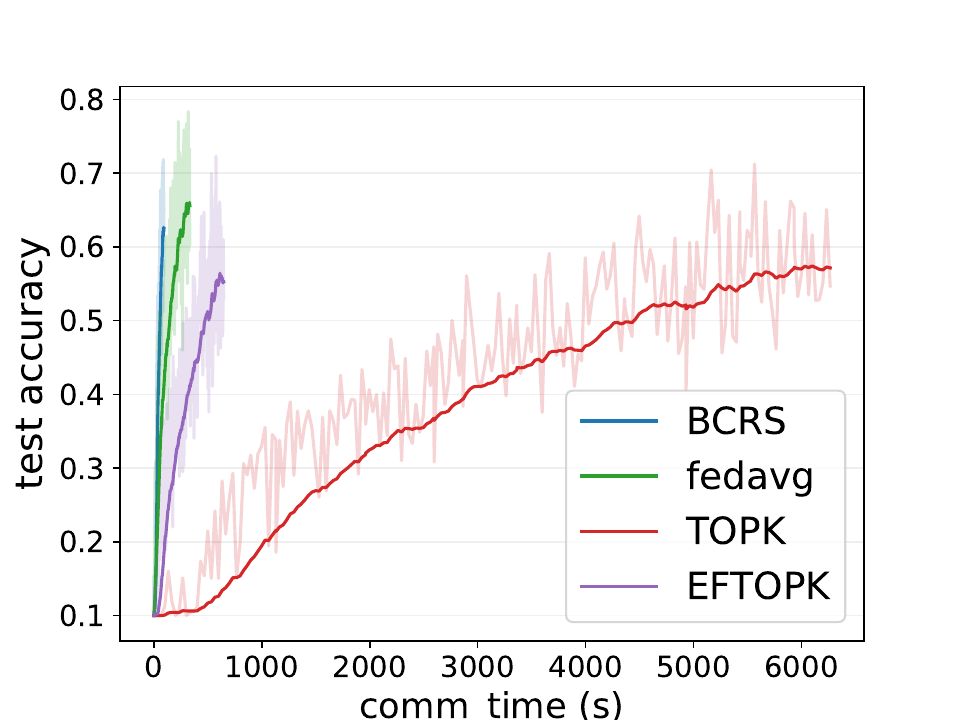}
	}
	\subfigure[$\beta$ = 0.1, CR = 0.01]
	{
	\includegraphics[width=0.46\linewidth]{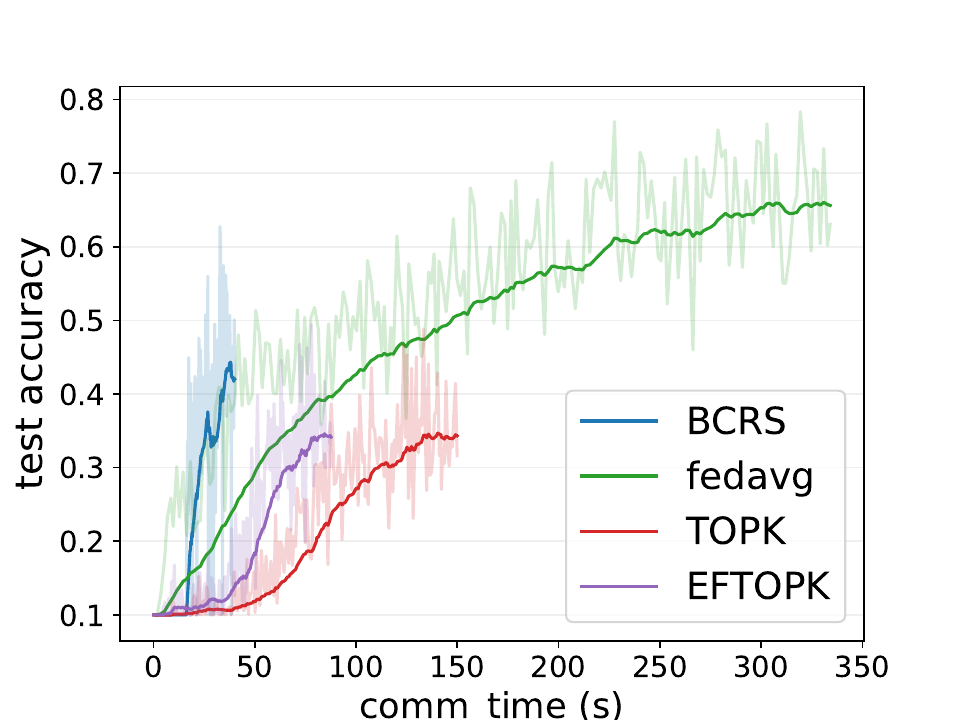}
	}
 \subfigure[$\beta$ = 0.5, CR= 0.1]
	{
	\includegraphics[width=0.46\linewidth]{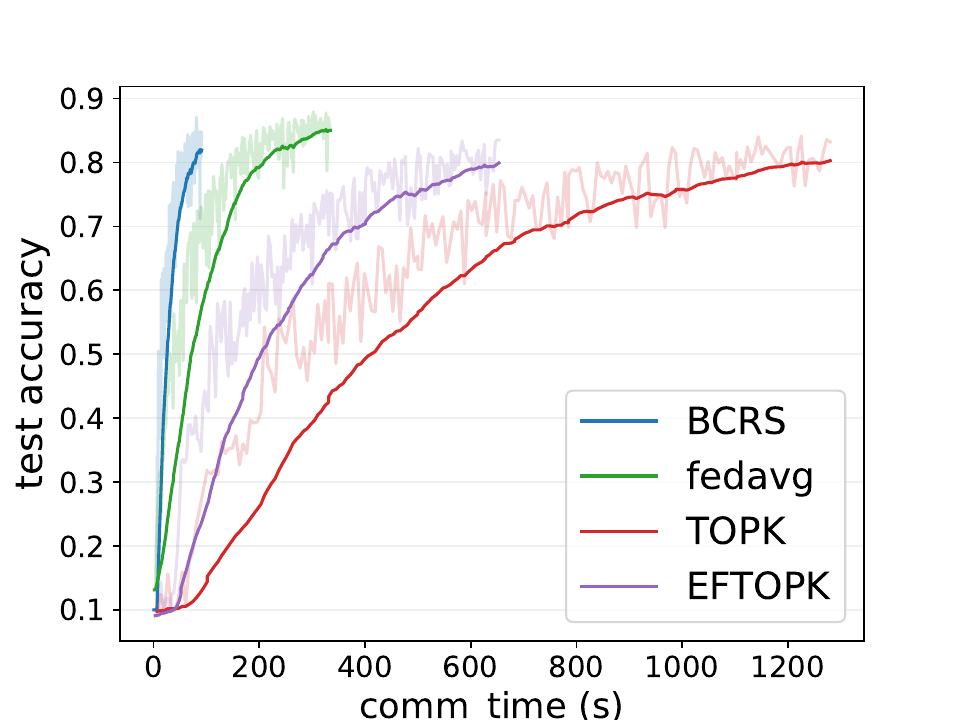}
	}
 \subfigure[$\beta$\ = 0.5\, CR = 0.01]
	{
	\includegraphics[width=0.46\linewidth]{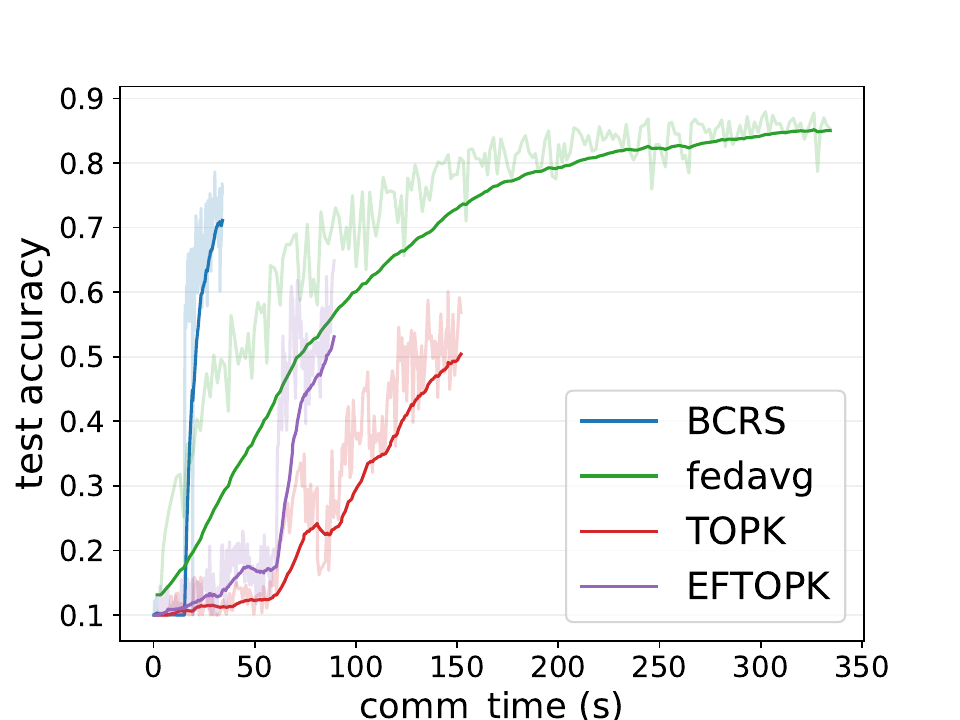}
	}
	\caption{CIFAR-10: Accuracy vs Communication time.}
	\label{fig:communicationeff}
\vspace{-0.25cm}
\end{figure}

It is imperative to underscore that the evaluation of all algorithms was simulated under random bandwidth and latency conditions. It is advisable to draw comparisons between the recorded Compressed Time and the corresponding accumulated Maximum time and Minimum time for each algorithm. Table~\ref{table:communication_efficiency} also reflects an intriguing balance between communication overhead and convergence rate. Employing a compression ratio does not directly translate to a proportional acceleration in the convergence rate. This finding further indicates the BCRS algorithm’s significant advantage in both ensuring accuracy and reducing the communication overhead.
Figure~\ref{fig:communicationeff} illustrates the relationship between accuracy and accumulated communication time under various settings. Notably, our BCRS algorithm demonstrates superior performance, achieving high accuracy with significantly reduced accumulated communication time compared to FedAvg and other baseline methods.

\subsubsection{Evaluation of Overlap-based Parameter Weighted Average}

\begin{figure}[!htb]
\setlength{\abovedisplayskip}{-1pt}
    \subfigbottomskip=0pt
    \subfigcapskip=0pt
    \setlength{\abovecaptionskip}{0.2cm}
\vspace{-0.4cm}
\centering
	\subfigure[$\beta=0.5$, CR=0.1]
	{
	\includegraphics[width=0.46\linewidth]{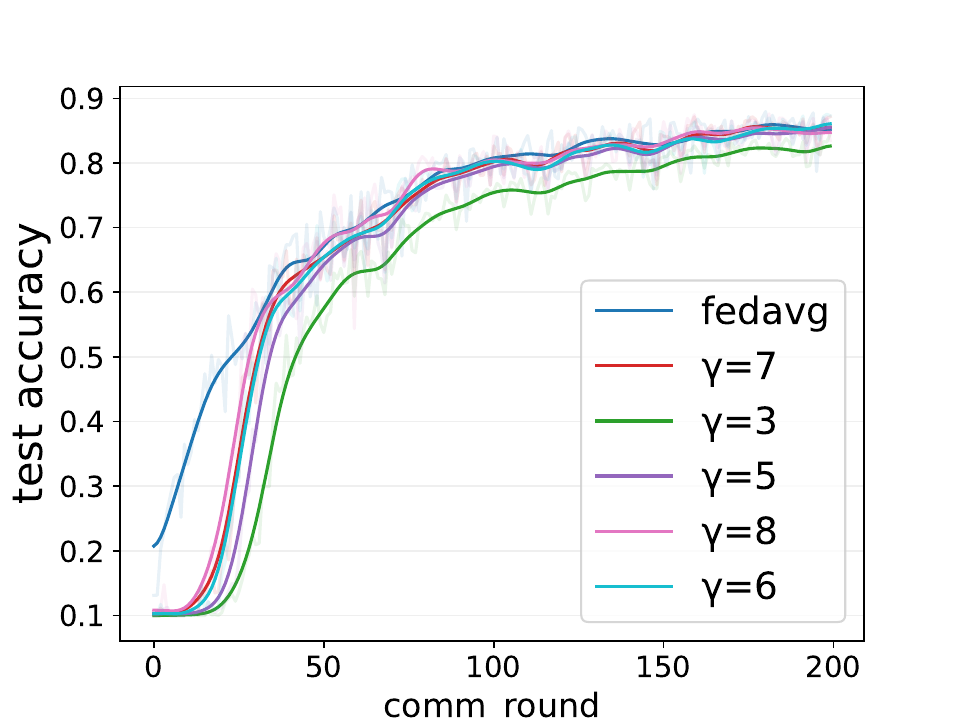}
	}
	\subfigure[$\beta$=0.1, CR=0.1]
	{
	\includegraphics[width=0.46\linewidth]{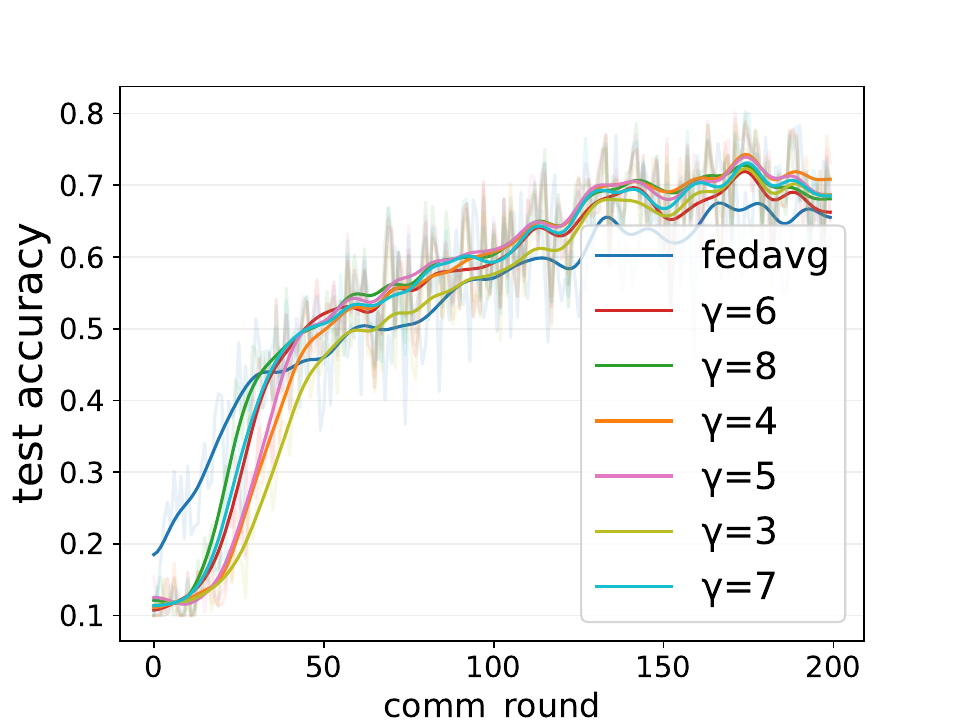}
	}
	\caption{CIFAR-10: Comparision of OPWA with different $\gamma$.}
	\label{fig:overlap_best_gamma_01}
\vspace{-0.2cm}
\end{figure}

In evaluating our OPWA algorithm, the focus is on the significant role played by the parameter enlarge rate $\gamma$. We explore a range of $\gamma$ values, from 1 up to the total number of clients ($N$), to gauge their effect on the algorithm's performance. The training curves for OPWA with different $\gamma$ configurations are shown in Figure.~\ref{fig:overlap_best_gamma_01}. We observe that the optimal $\gamma$ value is not necessarily within a certain candidate range. To better illustrate this observation, we select three representing values of $\gamma$, i.e. $\gamma=3, 5, 7$, and summarize the recorded accuracies and FedAvg's accuracy in Table~\ref{table:diff_beta}.

\textbf{Optimal $\gamma$ selection.} We can observe in Table~\ref{table:diff_beta} that the optimal enlarge rate $\gamma$ is not confined to the range [1, $\left| S_t\right|$] (which corresponds to the number of clients selected). When the best value of $\gamma$ falls in [$1$, $\left| S_t\right|$], we interpret it as the scenario where the updates of parameters with a low degree of overlap are insufficient due to the averaging process. On the other hand, the best enlarge rate $\gamma$ falling in the range [$\left| S_t\right| + 1$, $N$] is intriguing. It may suggest a balance between finding $\gamma$ and the optimal server learning rate $\alpha$ and the learning rate $\eta$. In this context, the larger enlarge rates compensate for $\alpha$ and $\eta$ that are not perfectly tuned for the FL environment.
\vspace{-0.2cm}
\begin{table}[H]
\centering
\fontsize{9}{11}\selectfont
\caption{OPWA test accuracy for different Enlarge Rates.}\label{table:diff_beta}    
    \begin{tabular}{|c|c|c||c|c|}
    \hline
         \multirow{2}{*}{Enlarge Rate $\gamma$} & \multicolumn{2}{c||}{$\beta=0.1$} &
         \multicolumn{2}{c|}{$\beta=0.5$}  \\\cline{2-5}
                    & CR=0.1 & CR=0.01 &  CR=0.1 & CR=0.01\\\hline\hline
        $\gamma=3$   &   0.5682 & 0.3461 & 0.6841 & 0.3282 \\ \hline
        $\gamma=5$  &   \textbf{0.5972} & 0.4222 & 0.7242 & 0.4809  \\ \hline
        $\gamma=7$  &  0.5958 & \textbf{0.4832} & \textbf{0.7375} & \textbf{0.5582} \\ \hline
        FedAvg  &  \multicolumn{2}{c|}{0.568} & \multicolumn{2}{c|}{0.7637} \\ \hline
    \end{tabular}
    \vspace{-0.5cm}
\end{table}

\begin{figure}[!htb]
\setlength{\abovedisplayskip}{-1pt}
    \subfigbottomskip=2pt
    \subfigcapskip=1pt
    \setlength{\abovecaptionskip}{0.2cm}
\vspace{-0.5cm}
\centering
	\subfigure[Clients $N = 16$]
	{
	\includegraphics[width=0.46\linewidth]{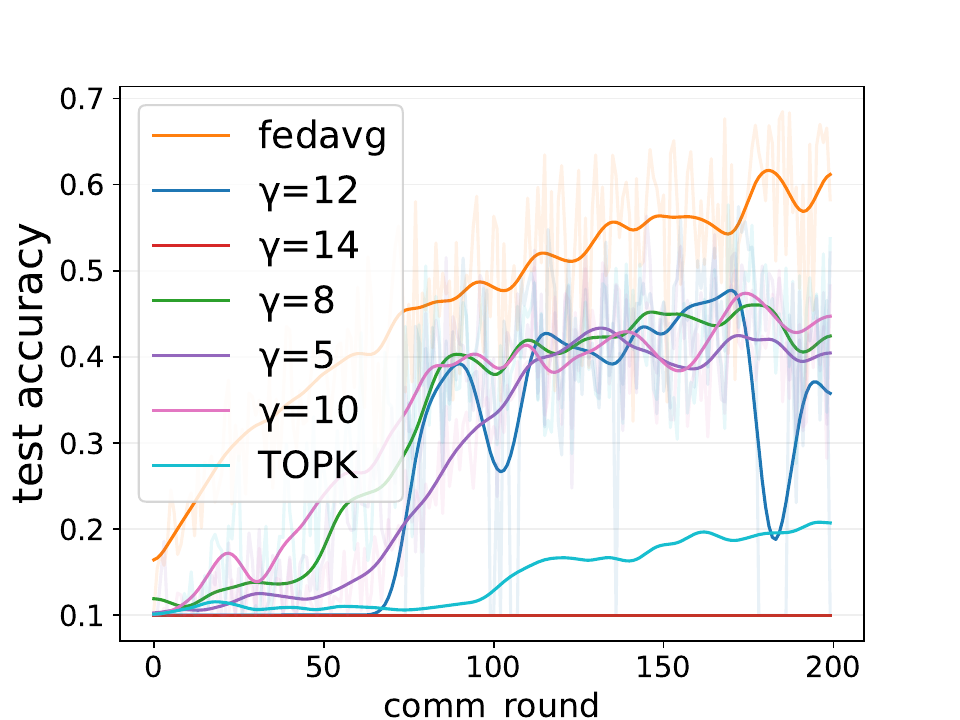}
	}
	\subfigure[Clients $N = 20$]
	{
	\includegraphics[width=0.46\linewidth]{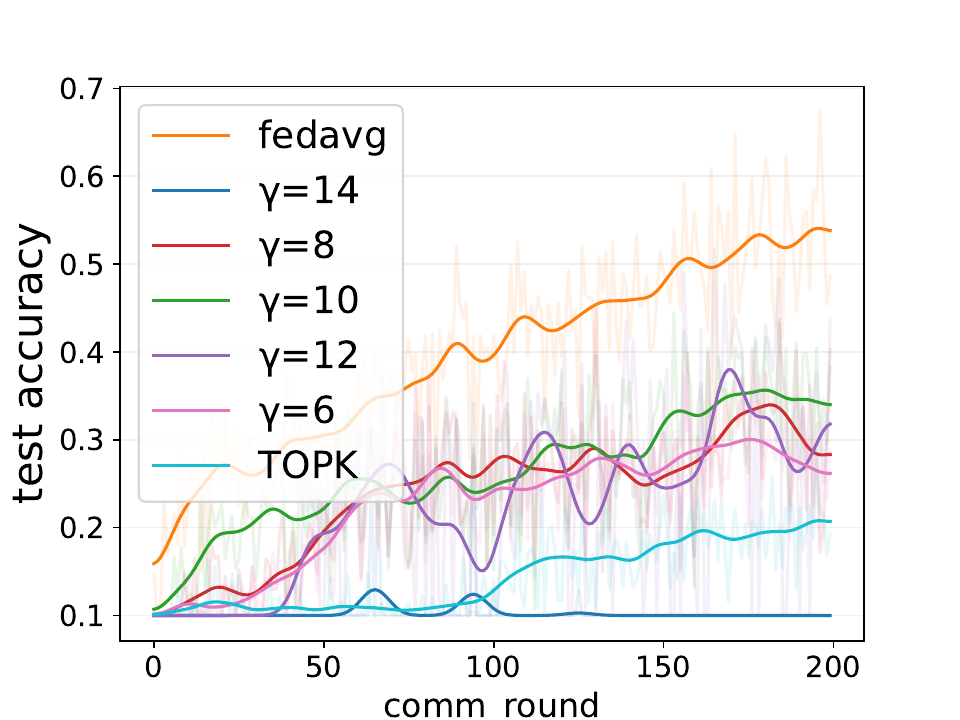}
	}
	\caption{CIFAR-10: Optimal $\gamma$ selection among different system scales ($N=16, 20$) with selection fraction 0.5.}
	\label{fig:overlap_best_gamma}
\vspace{-0.2cm}
\end{figure}

We scale up the number of clients participating in the training to test the selection of the optimal gamma on varying client counts. From Fig~\ref{fig:overlap_best_gamma}, We conclude that the optimal gamma is approximately proportional to the number of clients selected, which reflects the underrepresentation of such parameters in the model averaging.

\begin{figure}[!htb]
\setlength{\abovedisplayskip}{-1pt}
    \subfigbottomskip=0pt
    \subfigcapskip=0pt
    \setlength{\abovecaptionskip}{0.2cm}
\vspace{-0.5cm}
\centering
	\subfigure[$\beta$ = 0.1, CR = 0.01]
	{
	\includegraphics[width=0.46\linewidth]{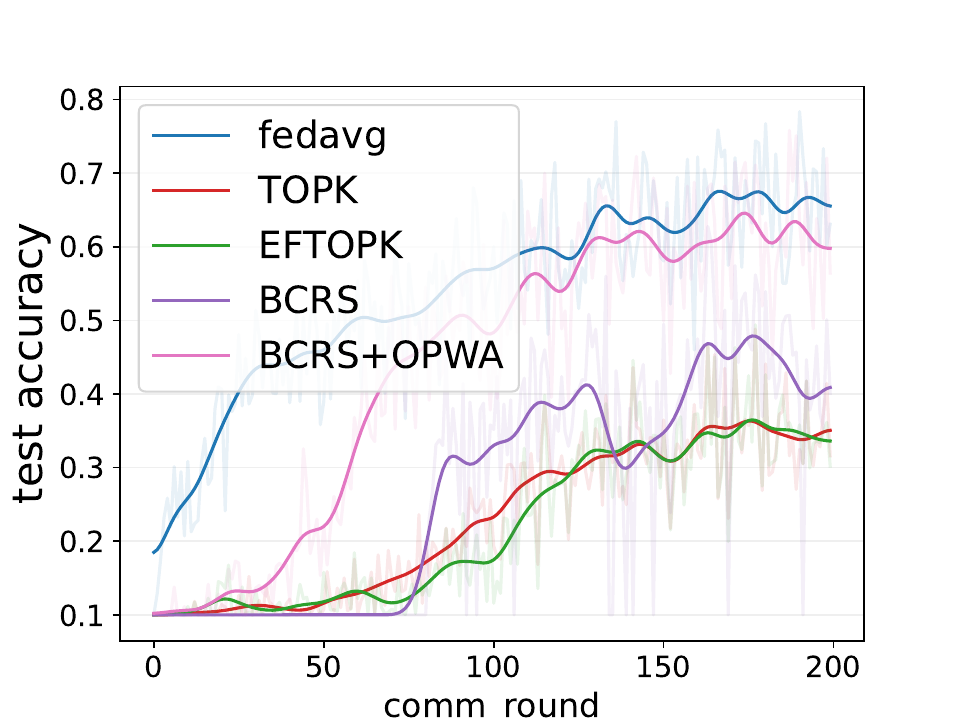}
	}
	\subfigure[$\beta$ = 0.1, CR = 0.1]
	{
	\includegraphics[width=0.46\linewidth]{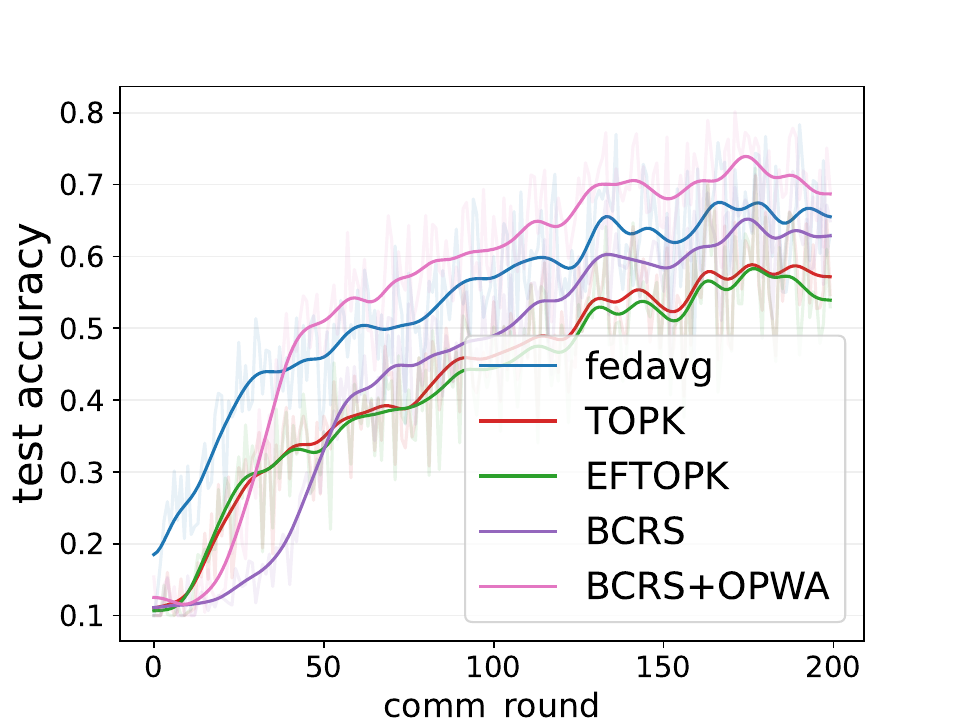}
	}
 \subfigure[$\beta$ = 0.5, CR = 0.1]
	{
	\includegraphics[width=0.46\linewidth]{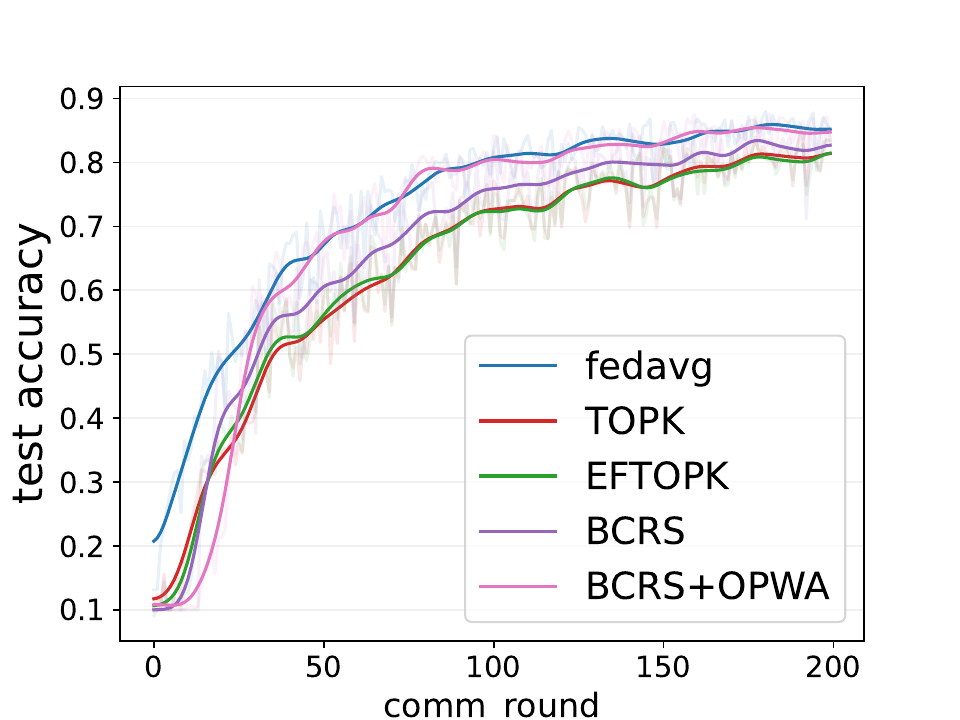}
	}
 \subfigure[$\beta$ = 0.5, CR = 0.01]
	{
	\includegraphics[width=0.46\linewidth]{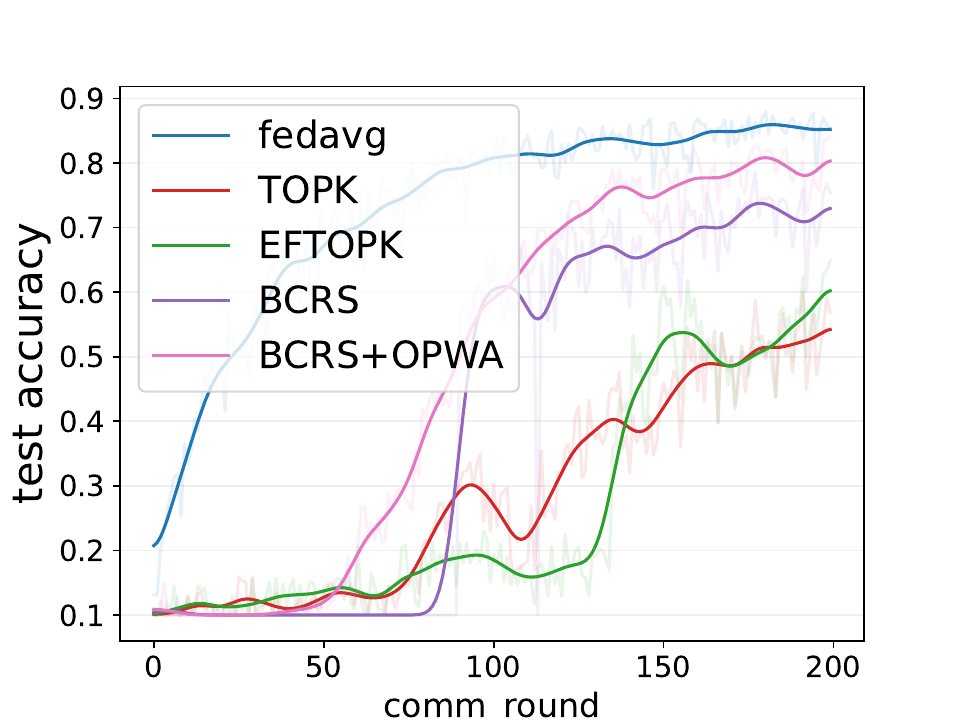}
	}
	\caption{CIFAR-10: Performance of OPWA and baselines.}
	\label{fig:overlap_best_7}
\vspace{-0.3cm}
\end{figure}

\begin{figure}[!h]
\setlength{\abovedisplayskip}{-1pt}
    \subfigbottomskip=2pt
    \subfigcapskip=1pt
    \setlength{\abovecaptionskip}{0.2cm}
  \centering
\vspace{-0.3cm}
\centering
\subfigure[$\beta$ = 0.1, CR = 0.1]
	{
	\includegraphics[width=0.46\linewidth]{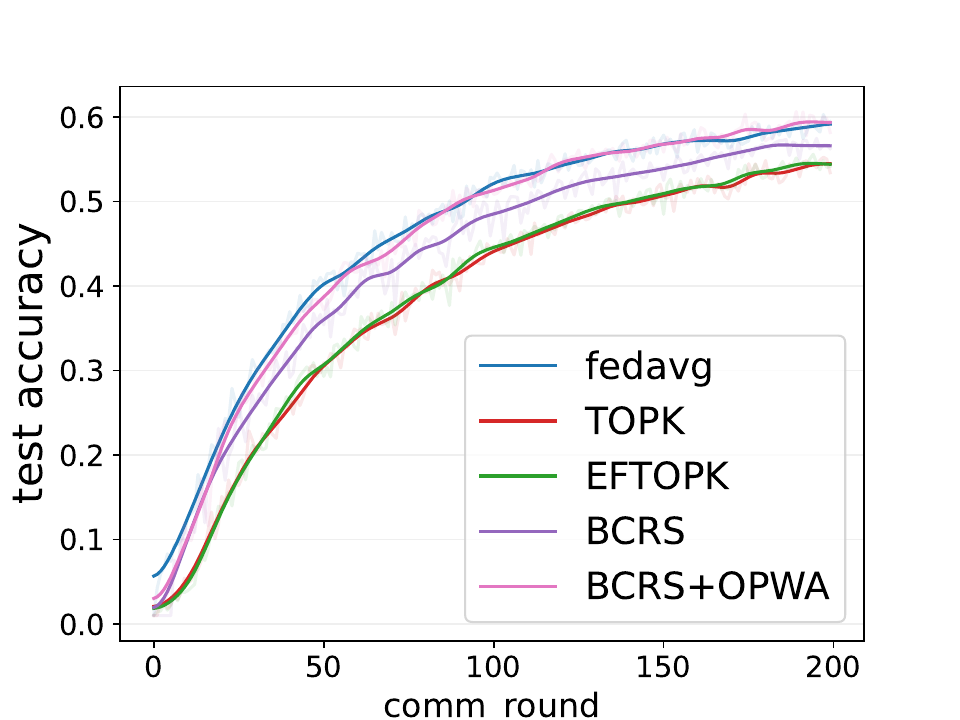}
	}
	\subfigure[$\beta$ = 0.1, CR = 0.01]
	{
	\includegraphics[width=0.46\linewidth]{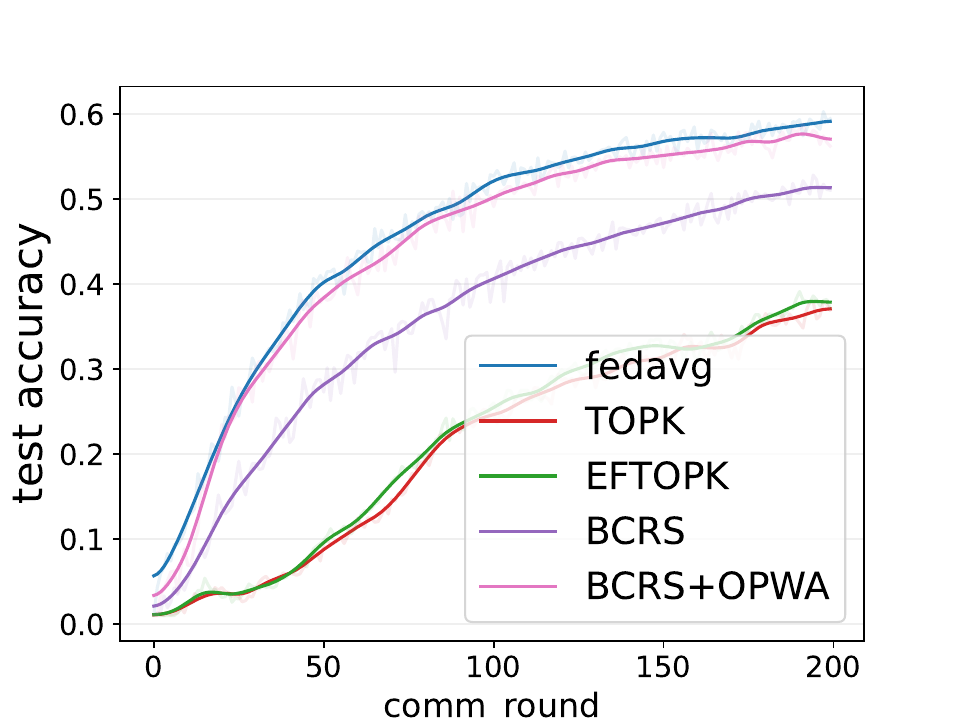}
	}
 \subfigure[$\beta$ = 0.5, CR= 0.1]
	{
	\includegraphics[width=0.46\linewidth]{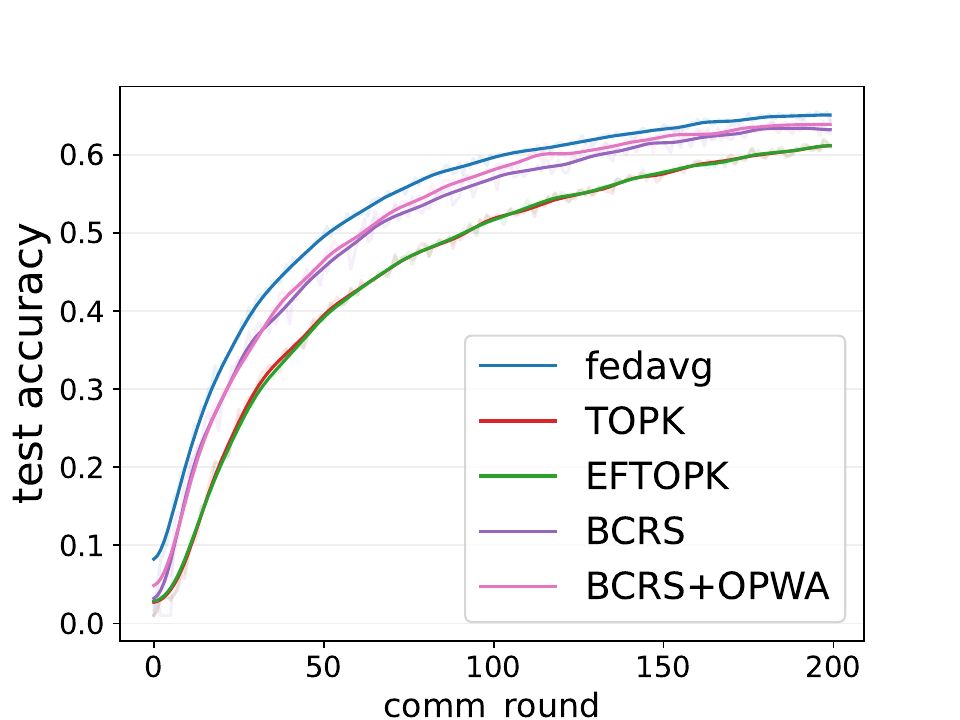}
	}
 \subfigure[$\beta$ = 0.5, CR = 0.01]
	{
	\includegraphics[width=0.46\linewidth]{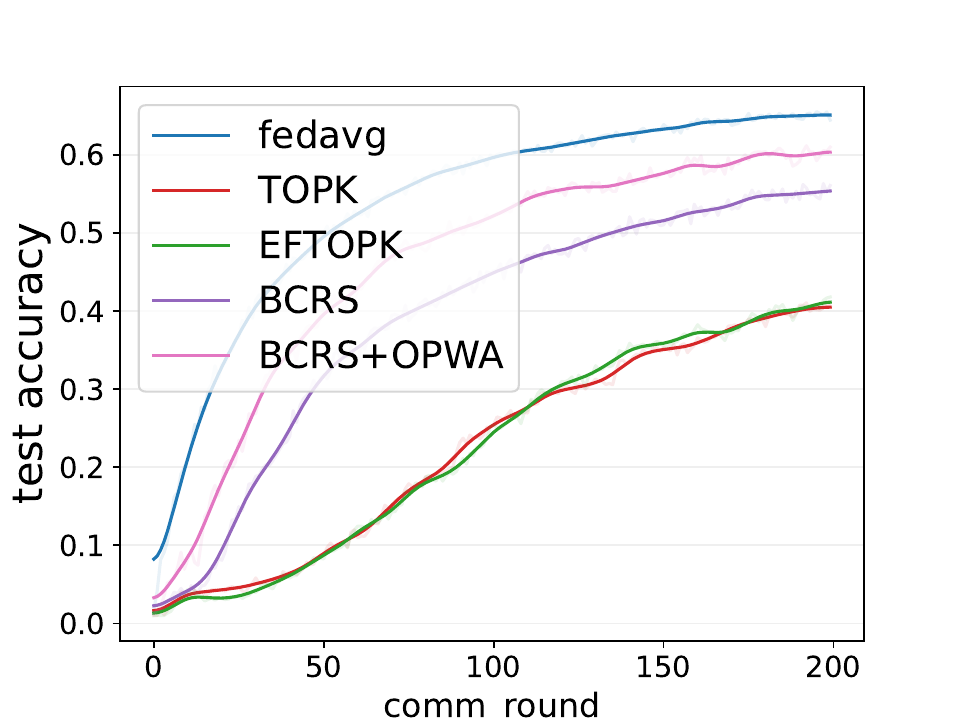}
	}
	\caption{CIFAR-100: OPWA and Baselines Comparison.}
	\label{fig:cifarhol}
\vspace{-0.4cm}
\end{figure}

\begin{figure}[!h]
\setlength{\abovedisplayskip}{-1pt}
    \subfigbottomskip=0pt
    \subfigcapskip=0pt
    \setlength{\abovecaptionskip}{0.2cm}
  \centering
\vspace{-0.3cm}
\centering
\subfigure[$\beta$ = 0.1, CR = 0.1]
	{
	\includegraphics[width=0.46\linewidth]{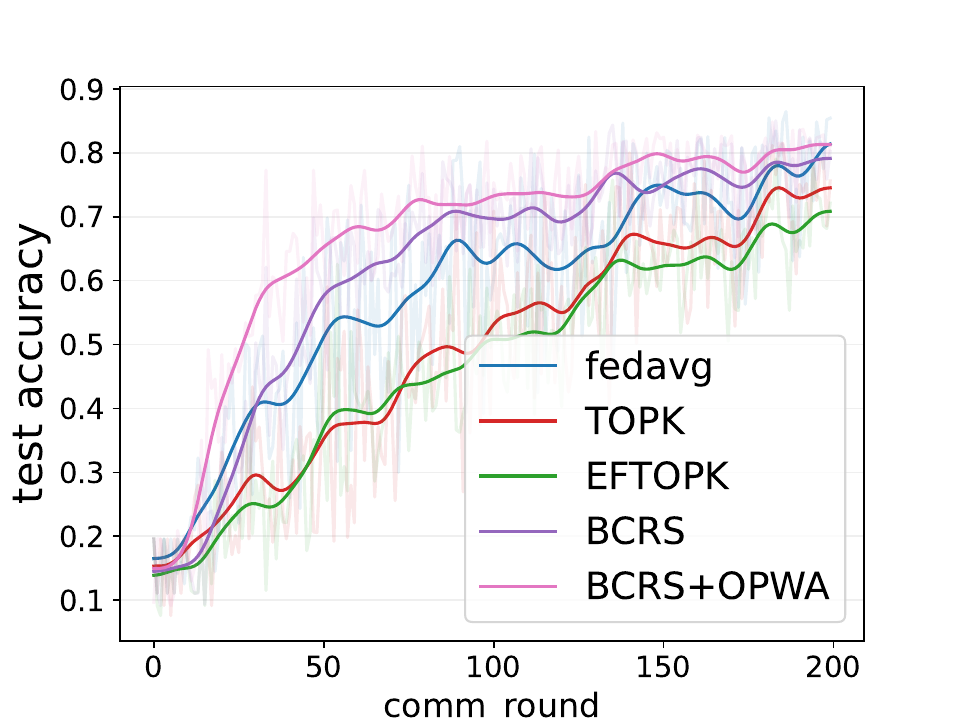}
	}
	\subfigure[$\beta$ = 0.1, CR = 0.01]
	{
	\includegraphics[width=0.46\linewidth]{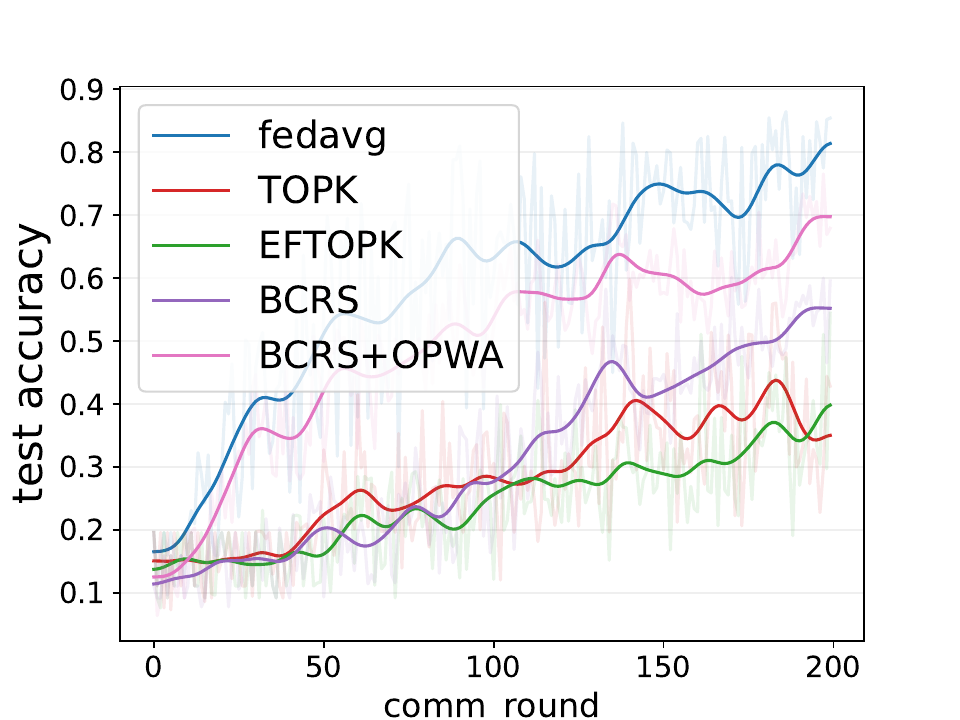}
	}
 \subfigure[$\beta$ = 0.5, CR= 0.1]
	{
	\includegraphics[width=0.46\linewidth]{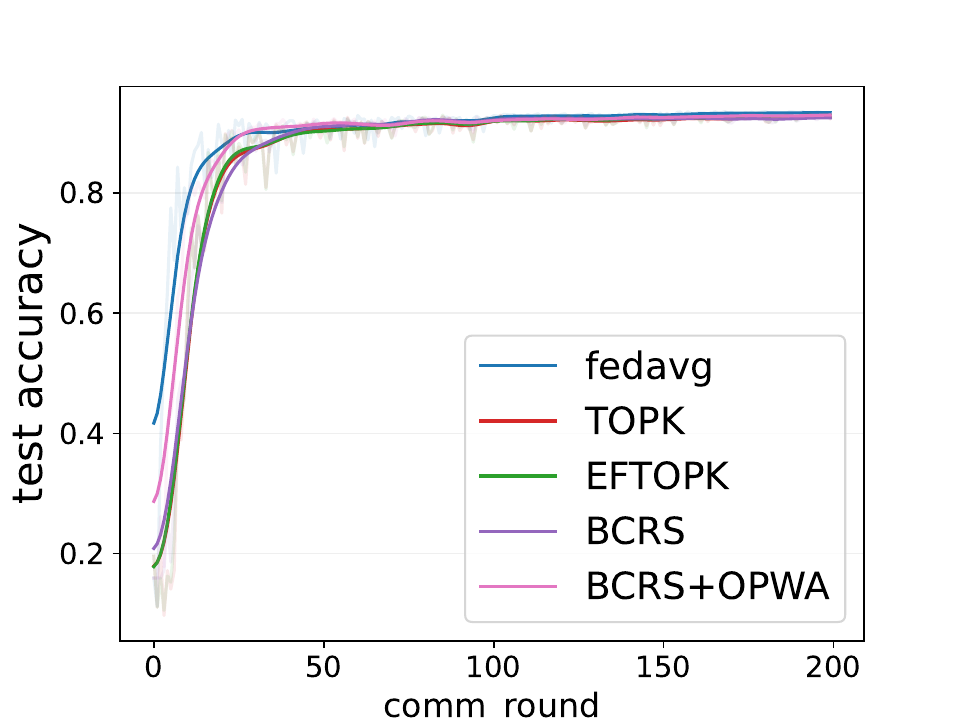}
	}
 \subfigure[$\beta$ = 0.5, CR = 0.01]
	{
	\includegraphics[width=0.46\linewidth]{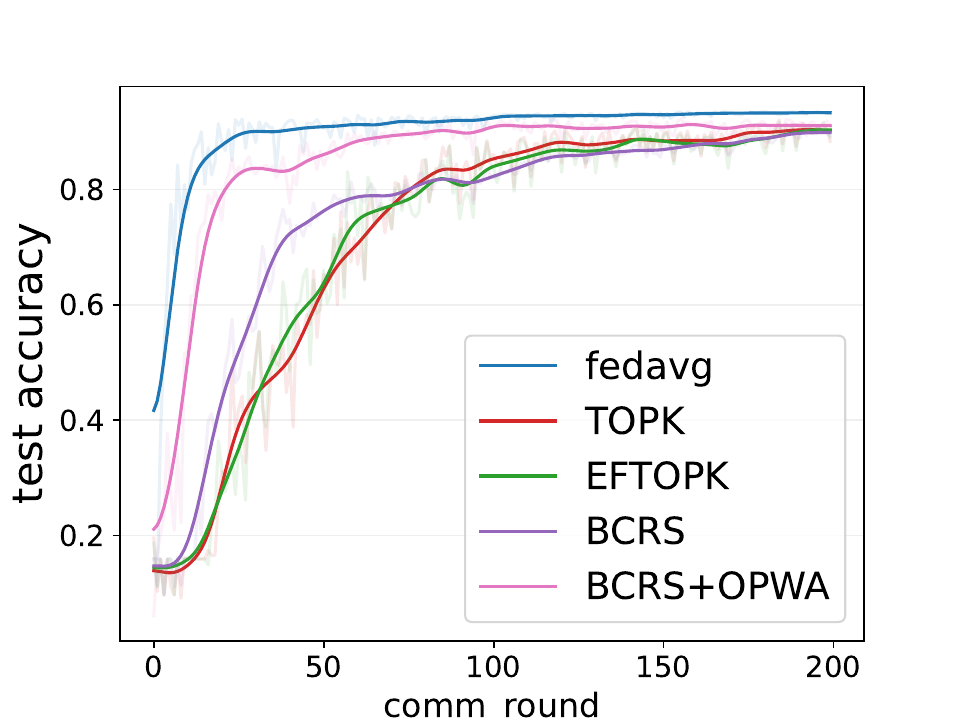}
	}
	\caption{SVHN: Comparison between OPWA and Baselines.}
	\label{fig:svhnol}
\vspace{-0.5cm}
\end{figure}

Furthermore, Fig.~\ref{fig:overlap_best_7} displays the training curves of both the baseline algorithms(FedAvg, TOP-K, EFTOPK, and BCRS) and the top-performing configuration of OPWA.
Under a compression ratio of 0.01, the OPWA algorithm demonstrated a substantial performance advantage over TOPK and EFTOK (approximately double the test accuracy). Remarkably, the performance of OPWA under this high compression ratio is comparable to that of the uncompressed FedAvg. We also observe in Fig.~\ref{fig:overlap_best_7} (b) that OPWA surpasses the performance of the uncompressed FedAvg at about round 60 under $CR=0.1$ and maintains its lead in test accuracy until the final round (round 200). These results highlight the effectiveness of OPWA in handling models with varying compression levels. The performance of OPWA on CIFAR-100 and SVHN are displayed in Fig~\ref{fig:cifarhol} and Fig~\ref{fig:svhnol}.



\section{Conclusion}
\label{Sec:conclusion}
In this work, we propose the BCRS framework that dynamically adjusts compression ratios and averaging coefficients based on bandwidth to solve the straggler problem caused by bandwidth heterogeneity. In addition, we find the non-overlap pattern of the retained parameters after compression and define a new metric to quantify the parameter overlap. Based on this observation, we design the OPWA technique to adjust the client-averaging weights at the parameter level to improve the convergence rate. This novel averaging technique can be incorporated seamlessly with other sparsification methods in FL. Furthermore, we conduct extensive experiments to demonstrate the improvement in the communication efficiency and model accuracy of the two proposed algorithms.

\begin{acks}
This work was partially supported by the National Natural Science Foundation of China under Grant No. 62272122, a Hong Kong RIF grant under Grant No. R6021-20, and Hong Kong CRF grants under Grant No. C2004-21G and C7004-22G.
\end{acks}
\bibliographystyle{ACM-Reference-Format}
\bibliography{cites.bib}

\appendix
\end{document}